\documentclass[tightenlines,superscriptaddress,10pt,twocolumn,aps,prd]{revtex4-2}

\usepackage{array}
\usepackage{lineno}
\usepackage{enumitem}
\usepackage{xspace}
\usepackage{placeins}
\usepackage{amsmath}    
\usepackage{graphicx}   
\usepackage{verbatim}   
\usepackage{color}      
\usepackage{subfigure}  
\usepackage{hyperref}   
\usepackage{silence}    

\newcommand{\ccpi}{CC$\pi^0$}
\newcommand{\ccpiid}{CC$\pi^0$ID}
\newcommand{\thereaction}{\ensuremath{\nu_\mu A\rightarrow\mu^-\pi^0 X}}

\begin{document}

\renewcommand{\thesection}{\arabic{section}}
\renewcommand{\thesubsection}{\thesection.\arabic{subsection}}
\renewcommand{\thesubsubsection}{\thesubsection.\arabic{subsubsection}}

\makeatletter
\renewcommand{\p@subsection}{}
\renewcommand{\p@subsubsection}{}
\makeatother


\begin{abstract}
Cross sections for the interaction \thereaction{} with neutrino energies between 1 and 5~GeV are measured using a sample of 165k selected events collected in the NOvA experiment's Near Detector, a hydrocarbon-based detector exposed to the NuMI neutrino beam at the Fermi National Accelerator Laboratory.  Results are presented as a flux-averaged total cross section and as differential cross sections in the momenta and angles of the outgoing muon and $\pi^0$, the total four-momentum transfer, and the invariant mass of the hadronic system.  Comparisons are made with predictions from a reference version of the GENIE neutrino interaction generator.  The measured total cross section of ($3.57\pm0.44)\times10^{-39}$\,cm$^2$ is $7.5\%$ higher than the GENIE prediction but is consistent within experimental errors.
\end{abstract}


\title{Measurement of $\nu_\mu$ Charged-Current Inclusive $\pi^0$ Production in the NOvA Near Detector}
\newcommand{\ANL}{Argonne National Laboratory, Argonne, Illinois 60439, 
USA}
\newcommand{\ICS}{Institute of Computer Science, The Czech 
Academy of Sciences, 
182 07 Prague, Czech Republic}
\newcommand{\IOP}{Institute of Physics, The Czech 
Academy of Sciences, 
182 21 Prague, Czech Republic}
\newcommand{\Atlantico}{Universidad del Atlantico,
Carrera 30 No. 8-49, Puerto Colombia, Atlantico, Colombia}
\newcommand{\BHU}{Department of Physics, Institute of Science, Banaras 
Hindu University, Varanasi, 221 005, India}
\newcommand{\UCLA}{Physics and Astronomy Department, UCLA, Box 951547, Los 
Angeles, California 90095-1547, USA}
\newcommand{\Caltech}{California Institute of 
Technology, Pasadena, California 91125, USA}
\newcommand{\Cochin}{Department of Physics, Cochin University
of Science and Technology, Kochi 682 022, India}
\newcommand{\Charles}
{Charles University, Faculty of Mathematics and Physics,
 Institute of Particle and Nuclear Physics, Prague, Czech Republic}
\newcommand{\Cincinnati}{Department of Physics, University of Cincinnati, 
Cincinnati, Ohio 45221, USA}
\newcommand{\CSU}{Department of Physics, Colorado 
State University, Fort Collins, CO 80523-1875, USA}
\newcommand{\CTU}{Czech Technical University in Prague,
Brehova 7, 115 19 Prague 1, Czech Republic}
\newcommand{\Dallas}{Physics Department, University of Texas at Dallas,
800 W. Campbell Rd. Richardson, Texas 75083-0688, USA}
\newcommand{\DallasU}{University of Dallas, 1845 E 
Northgate Drive, Irving, Texas 75062 USA}
\newcommand{\Delhi}{Department of Physics and Astrophysics, University of 
Delhi, Delhi 110007, India}
\newcommand{\JINR}{Joint Institute for Nuclear Research,  
Dubna, Moscow region 141980, Russia}
\newcommand{\FNAL}{Fermi National Accelerator Laboratory, Batavia, 
Illinois 60510, USA}
\newcommand{\UFG}{Instituto de F\'{i}sica, Universidade Federal de 
Goi\'{a}s, Goi\^{a}nia, Goi\'{a}s, 74690-900, Brazil}
\newcommand{\Guwahati}{Department of Physics, IIT Guwahati, Guwahati, 781 
039, India}
\newcommand{\Harvard}{Department of Physics, Harvard University, 
Cambridge, Massachusetts 02138, USA}
\newcommand{\Houston}{Department of Physics, 
University of Houston, Houston, Texas 77204, USA}
\newcommand{\IHyderabad}{Department of Physics, IIT Hyderabad, Hyderabad, 
502 205, India}
\newcommand{\Hyderabad}{School of Physics, University of Hyderabad, 
Hyderabad, 500 046, India}
\newcommand{\IIT}{Department of Physics,
Illinois Institute of Technology,
Chicago IL 60616, USA}
\newcommand{\Indiana}{Indiana University, Bloomington, Indiana 47405, 
USA}
\newcommand{\INR}{Institute for Nuclear Research of Russia, Academy of 
Sciences 7a, 60th October Anniversary prospect, Moscow 117312, Russia}
\newcommand{\Iowa}{Department of Physics and Astronomy, Iowa State 
University, Ames, Iowa 50011, USA}
\newcommand{\Irvine}{Department of Physics and Astronomy, 
University of California at Irvine, Irvine, California 92697, USA}
\newcommand{\Jammu}{Department of Physics and Electronics, University of 
Jammu, Jammu Tawi, 180 006, Jammu and Kashmir, India}
\newcommand{\Lebedev}{Nuclear Physics and Astrophysics Division, Lebedev 
Physical 
Institute, Leninsky Prospect 53, 119991 Moscow, Russia}
\newcommand{\Magdalena}{Universidad del Magdalena, Carrera 32 No 22 – 08 
Santa Marta, Colombia}
\newcommand{\MSU}{Department of Physics and Astronomy, Michigan State 
University, East Lansing, Michigan 48824, USA}
\newcommand{\Crookston}{Math, Science and Technology Department, University 
of Minnesota Crookston, Crookston, Minnesota 56716, USA}
\newcommand{\Duluth}{Department of Physics and Astronomy, 
University of Minnesota Duluth, Duluth, Minnesota 55812, USA}
\newcommand{\Minnesota}{School of Physics and Astronomy, University of 
Minnesota Twin Cities, Minneapolis, Minnesota 55455, USA}
\newcommand{\Oxford}{Subdepartment of Particle Physics, 
University of Oxford, Oxford OX1 3RH, United Kingdom}
\newcommand{\Panjab}{Department of Physics, Panjab University, 
Chandigarh, 160 014, India}
\newcommand{\Pitt}{Department of Physics, 
University of Pittsburgh, Pittsburgh, Pennsylvania 15260, USA}
\newcommand{\RAL}{Rutherford Appleton Laboratory, Science and 
Technology Facilities Council, Didcot, OX11 0QX, United Kingdom}
\newcommand{\SAlabama}{Department of Physics, University of 
South Alabama, Mobile, Alabama 36688, USA} 
\newcommand{\Carolina}{Department of Physics and Astronomy, University of 
South Carolina, Columbia, South Carolina 29208, USA}
\newcommand{\SDakota}{South Dakota School of Mines and Technology, Rapid 
City, South Dakota 57701, USA}
\newcommand{\SMU}{Department of Physics, Southern Methodist University, 
Dallas, Texas 75275, USA}
\newcommand{\Stanford}{Department of Physics, Stanford University, 
Stanford, California 94305, USA}
\newcommand{\Sussex}{Department of Physics and Astronomy, University of 
Sussex, Falmer, Brighton BN1 9QH, United Kingdom}
\newcommand{\Syracuse}{Department of Physics, Syracuse University,
Syracuse NY 13210, USA}
\newcommand{\Tennessee}{Department of Physics and Astronomy, 
University of Tennessee, Knoxville, Tennessee 37996, USA}
\newcommand{\Texas}{Department of Physics, University of Texas at Austin, 
Austin, Texas 78712, USA}
\newcommand{\Tufts}{Department of Physics and Astronomy, Tufts University, Medford, 
Massachusetts 02155, USA}
\newcommand{\UCL}{Physics and Astronomy Dept., University College London, 
Gower Street, London WC1E 6BT, United Kingdom}
\newcommand{\Virginia}{Department of Physics, University of Virginia, 
Charlottesville, Virginia 22904, USA}
\newcommand{\WSU}{Department of Mathematics, Statistics, and Physics,
 Wichita State University, 
Wichita, Kansas 67206, USA}
\newcommand{\WandM}{Department of Physics, William \& Mary, 
Williamsburg, Virginia 23187, USA}
\newcommand{\Winona}{Department of Physics, Winona State University, P.O. 
Box 5838, Winona, Minnesota 55987, USA}
\newcommand{\Wisconsin}{Department of Physics, University of 
Wisconsin-Madison, Madison, Wisconsin 53706, USA}
\newcommand{\deceased}{Deceased.}
\affiliation{\ANL}
\affiliation{\Atlantico}
\affiliation{\BHU}
\affiliation{\Caltech}
\affiliation{\Charles}
\affiliation{\Cincinnati}
\affiliation{\Cochin}
\affiliation{\CSU}
\affiliation{\CTU}
\affiliation{\DallasU}
\affiliation{\Delhi}
\affiliation{\FNAL}
\affiliation{\UFG}
\affiliation{\Guwahati}
\affiliation{\Harvard}
\affiliation{\Houston}
\affiliation{\Hyderabad}
\affiliation{\IHyderabad}
\affiliation{\Indiana}
\affiliation{\ICS}
\affiliation{\IIT}
\affiliation{\INR}
\affiliation{\IOP}
\affiliation{\Iowa}
\affiliation{\Irvine}
\affiliation{\Jammu}
\affiliation{\JINR}
\affiliation{\Lebedev}
\affiliation{\MSU}
\affiliation{\Duluth}
\affiliation{\Magdalena}
\affiliation{\Minnesota}
\affiliation{\Panjab}
\affiliation{\Pitt}
\affiliation{\SAlabama}
\affiliation{\Carolina}
\affiliation{\SDakota}
\affiliation{\SMU}
\affiliation{\Stanford}
\affiliation{\Sussex}
\affiliation{\Syracuse}
\affiliation{\Tennessee}
\affiliation{\Texas}
\affiliation{\Tufts}
\affiliation{\UCL}
\affiliation{\Virginia}
\affiliation{\WSU}
\affiliation{\WandM}
\affiliation{\Wisconsin}

\author{M.~A.~Acero}
\affiliation{\Atlantico}

\author{P.~Adamson}
\affiliation{\FNAL}

\author{G.~Agam}
\affiliation{\IIT}


\author{L.~Aliaga}
\affiliation{\FNAL}

\author{T.~Alion}
\affiliation{\Sussex}

\author{V.~Allakhverdian}
\affiliation{\JINR}

\author{S.~Altakarli}
\affiliation{\WSU}



\author{N.~Anfimov}
\affiliation{\JINR}


\author{A.~Antoshkin}
\affiliation{\JINR}

\author{L.~Asquith}
\affiliation{\Sussex}


% ...

\collaboration{The NOvA Collaboration}
\noaffiliation

\date{\today}

\maketitle

\section{Introduction}

Muon-neutrino-induced charged-current (CC) $\pi^0$ production on a nuclear target, hereinafter ``\ccpi{},'' is the reaction
\begin{equation}
\thereaction{}\ ,
\end{equation}
where $A$ is the target nucleus and $X$ represents the final state nucleus plus any additional reaction products, possibly including other charged or neutral pions.  This channel is of particular interest for experiments studying $\nu_\mu\rightarrow\nu_e$ flavor oscillations not only because it can directly lead to backgrounds in those measurements but also because of its close relation to the important background process of neutral current $\pi^0$ production.  Events in which only one photon from the $\pi^0$ decay is reconstructed can be mistaken as containing a primary electron, the defining characteristic of a $\nu_e$ CC event.

Improved understanding of neutrino-nucleus interactions is of great benefit for the current and next generation of neutrino oscillation experiments.  Experiments such as NOvA \cite{bib:NOvAJointFit} and DUNE \cite{bib:DUNETDR} lie in the few-GeV transition region between the low energy regime dominated by quasi-elastic (QE) scattering and the high energy regime dominated by deep inelastic scattering (DIS).  In addition to QE and DIS events, this transition region has a large component of baryon resonance (Res) events in which a neutrino scatters off a nucleon producing an intermediate $\Delta(1232)$ or higher-mass baryon.  Pions are commonly produced in both DIS and resonant baryon interactions but through very different production mechanisms.  The picture is complicated further in a nuclear medium by final-state interactions (FSI), in which outgoing hadronic particles may strongly interact with nucleons before escaping the target nucleus.  Experiments are sensitive only to particles exiting the nucleus, making it difficult to attribute observed discrepancies with a model to either the neutrino interaction or subsequent FSI effects.

The first measurements of neutrino-induced $\pi^0$ production were exclusive analyses of bubble chamber data~\cite{bib:ANLCCPi0, bib:BNLCCPi0}.  Exclusive or  semi-inclusive cross sections on hydrocarbon targets have been studied in MINERvA~\cite{bib:MINERvACCPi0-2015, bib:MINERvACCPi0-2017, PhysRevD.102.072007}, MiniBooNE~\cite{bib:MiniBooNECCPi0} and K2K \cite{bib:K2KCCPi0}.  Charged-current $\pi^0$ production on argon has also been recently measured in MicroBooNE~\cite{PhysRevD.99.091102}.  The present analysis examines fully inclusive $\pi^0$ production on a primarily hydrocarbon target.

We present measurements of the \ccpi{} flux-averaged total cross section as well as cross sections differential in $p_\pi$; $\cos\theta_\pi$; $p_\mu$; $\cos\theta_\mu$; lepton momentum transfer $Q^2$ as defined below in Eq.~(\ref{eqn:Q2}); and the invariant mass $W$ of the hadronic system as defined below in Eq.~(\ref{eqn:W}).  The data used for these measurements was collected between August 2014 and January 2016 and corresponds to an exposure of 3.72$\times10^{20}$ protons-on-target (POT) using a predominantly $\nu_\mu$ beam. 165k events are selected as $\nu_\mu$~CC candidates, of which simulation predicts $34\%$ are signal \ccpi{}.  A subsequent fit to a \ccpi{} event classifier variable provides a final stage of background separation and signal rate estimation.  With this large data set, the uncertainty is dominated by systematic sources.

\section{The NOvA Experiment}

The NOvA experiment \cite{bib:NOvAJointFit} is designed to measure neutrino oscillations over an 810\,km baseline.  It uses two functionally identical detectors situated along the Neutrinos at the Main Injector (NuMI) beam from the Fermi National Accelerator Laboratory (FNAL).  The NOvA near detector (ND) is located 1\,km downstream of the beam target where it is subject to an intense neutrino flux.  The ND has recorded millions of neutrino and antineutrino interactions, allowing large-sample neutrino cross-section measurements.

\subsection{The NuMI Beam}

The NuMI neutrino beam \cite{bib:NuMI} is created by directing 120\,GeV protons from the FNAL Main Injector onto a graphite target.  The many charged pions and kaons produced in the collision are focused through two magnetic horns and allowed to decay in a 650 m decay pipe to produce the primarily $\nu_\mu$ neutrino beam.  The NOvA detectors are located 14.6\,mrad off-axis relative to the beam centerline, resulting in a narrow-band flux peaked at $E_\nu = 1.8$\,GeV.  There are two classes of impurity in the neutrino flux in the neutrino-dominated beam configuration: $\bar{\nu}_\mu$ which accounts for 1.8$\%$ of neutrinos in the 1--3\,GeV region around the beam peak and $\nu_e+\bar{\nu}_e$ which accounts for 0.7$\%$ in the same energy range.

\subsection{The NOvA Near Detector}

The ND is a tracking calorimeter with fine segmentation relative to the 40\,cm radiation length for precise imaging of electromagnetic showers.  The detector is built of PVC cells with a 3.9\,cm (transverse) by 6.6\,cm (longitudinal) cross section and a length of 3.9\,m.    
A detector plane consists of 96 cells, and planes are arranged in alternating vertical and horizontal orientations, allowing  3D reconstruction of observed events.
The fully active volume of the detector is 12.8\,m in length, consisting of 192 contiguous  planes. 
Each detector cell is filled with a liquid scintillator blend that is 
95$\%$ mineral oil and 
5$\%$ pseudocumene with trace concentrations of wavelength shifting fluors~\cite{bib:Scintillator}.  Each cell contains a wavelength shifting fiber that collects and delivers light to an avalanche photodiode.

Additionally, a muon range stack is situated at the downstream end of the detector, consisting of 11 pairs of readout planes with a layer of 10.2-cm-thick steel between adjacent pairs. 
The range stack increases the muon energy that can be contained in the detector to about 4.5\,GeV.

The interaction fiducial volume does not include any part of the range stack, limiting the relevant nuclear targets to those in the PVC and scintillator.  A precise accounting of the scattering material is given in Sec.~\ref{sect:Ana:TargetCount}.

\section{Event Simulation}

Predicted event rates are calculated with a detailed simulation in three stages: production and transport of the neutrino beam, neutrino interaction, and detector response.

\subsection{Neutrino Beam Simulation}
\label{sec:BeamSim}
Simulated outgoing hadrons from proton-nucleus collisions within the NuMI target are modeled with FLUKA~\cite{bib:FLUKA2, bib:FLUKA1}. The charged pions and kaons are subsequently propagated via FLUGG~\cite{bib:FLUGG} through the focusing horns and into the decay pipe until their decay.

The neutrino flux prediction is improved according to the PPFX \cite{bib:MINERvAPPFX} framework. PPFX combines hadron production data from an extensive survey of proton-nucleus scattering experiments, and is used to constrain the predicted hadron multiplicities exiting the target.  Both the central value and error band from the PPFX prediction are used.  The central value for the neutrino flux integrated between 1 and 5\,GeV after PPFX corrections is 8.2\% lower than the raw FLUKA and FLUGG prediction. 

\subsection{Neutrino Interaction Simulation}

Neutrino interactions are simulated with GENIE~\cite{bib:GENIE1} v2.10.2.  The GENIE simulation generates interactions via its four default production processes: quasi-elastic scattering, resonant baryon production, deep-inelastic scattering, and coherent pion production. Particles creating via these primary processes are subsequently propagated though the nuclear medium using GENIE's {\em hA} effective cascade final-state interaction (FSI) model~\cite{Dytman:2007zz,Dytman:2009zz}.

For quasi-elastic interactions, GENIE uses a Llewellyn Smith parameterization \cite{bib:LlewellynSmith}.  Resonant baryon production follows the Rein-Sehgal model \cite{bib:ResReinSeghal}, which calculates the resonant cross section through production and decay of $\Delta(1232)$ and higher mass $N^*$ baryon resonances.  Coherent pion production interactions are also simulated with the Rein-Sehgal model. Deep-inelastic scattering interactions are generated according to the Bodek-Yang model \cite{bib:BodekYang}.

Meson exchange current (MEC) interactions are insignificant in this measurement given the signal selection, which requires evidence of a neutral pion.  So that GENIE can be used in its documented default configuration and to avoid concerns over certain specifics of early MEC models in this GENIE version, MEC events were not enabled in the baseline simulation.  Note that the similar quasi-elastic channel represents 0.4\% of the selected sample.  The negligible impact of MEC events on the reported cross sections was confirmed using a separate simulated event sample based on the empirical Dytman MEC model \cite{bib:MECModels}.

\subsection{Detector Response Simulation}

Final state particles produced by GENIE are propagated through the detector using GEANT~4.9.6.p04d \cite{bib:GEANT4}.  Optical photon production is modeled with the Birks-Chou parameterization \cite{bib:BirksChou} for scintillator response.  The light collection, signal transport in the fibers, and photodiode and electronics response are modeled using custom simulation software \cite{bib:CHEP2015_sim}.  The detector is calibrated using minimum ionizing portions of stopping cosmic ray muon tracks.  The detector response to hadronic tracks and electromagnetic showers is tested using two control samples, each discussed in detail in Sec.~\ref{syst:DetResponse}.

\section{Signal Definition}

For this analysis a \ccpi{} event is defined as any $\nu_\mu$ CC event with at least one $\pi^0$ emerging from the struck nucleus.  This definition includes events with multiple neutral or charged pions in the final state.  Multi-$\pi$ events are common at NOvA energies, representing $54\%$ of \ccpi{} events in this analysis and also accounting for the majority of NC and $\nu_\mu$-induced CC backgrounds in the NOvA $\nu_\mu\rightarrow\nu_e$ oscillation measurement~\cite{bib:NOvAJointFit}.
 
Signal events are further required to lie within the kinematic region specified in Table~\ref{table:KinRegion}.  The table also gives the rationale for the kinematic exclusions applied. Selected \ccpi{} events whose true kinematics lie outside these ranges but that leak into the selected sample are treated as an analysis background and comprise only 0.2\% of the sample.

\begin{table}[tb!]
\caption{Definition of the signal region for \ccpi{} interactions. 
The left column specifies kinematic regions excluded from the signal definition while the right lists the motivation for excluding events that lie in the excluded regions.}

\centering
\begin{tabular}{ c | c }
Kinematic Exclusion & Motivation \\
\hline
$p_\pi > 3\,\mathrm{GeV}/c$ & Negligible rate \\
$p_\mu > 4\,\mathrm{GeV}/c$ & Long muons uncontained \\
$E_\nu < 1$\,GeV & Background dominated \\
$E_\nu > 5$\,GeV & Imprecise flux modeling \\
$Q^2 > 3\,\mathrm{GeV}^2/c^2$ & Negligible rate \\
$W < 1\,\mathrm{GeV}/c^2$ & Background dominated \\
$W > 3\,\mathrm{GeV}/c^2$ & Negligible rate
\end{tabular}

\label{table:KinRegion}
\end{table}

\section{Event Reconstruction and Selection}

\begin{figure*}
\centering
\includegraphics[width=0.95\linewidth]{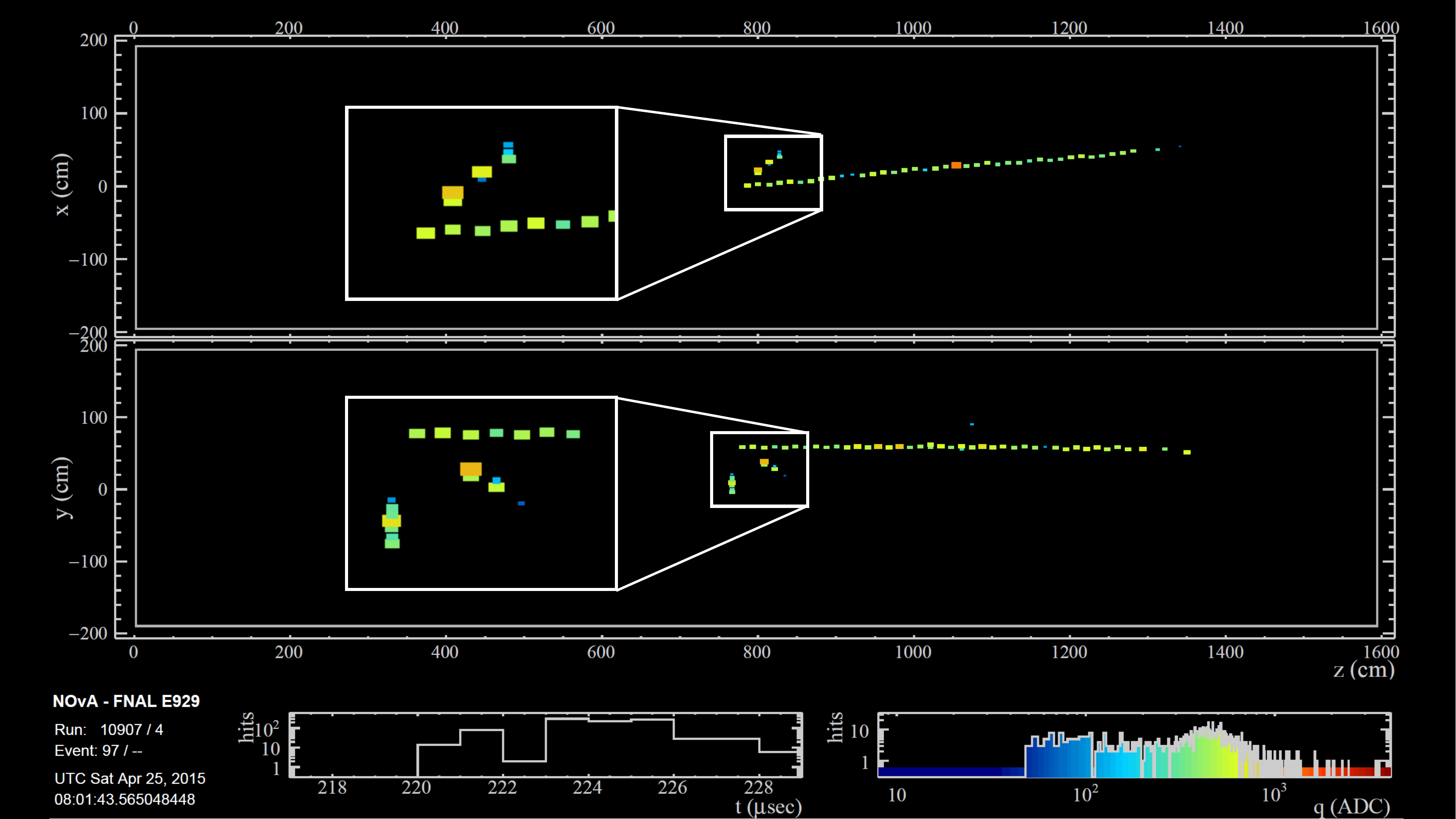}
\caption{An example simulated ND event.  The top panel shows the overhead view of the particle tracks in the detector while the bottom shows the side view of the same tracks.  In each view, a small white box shows the vertex region, with a magnified version shown in a large white box. Individual detector cells with recorded energy depositions are shown as colored rectangles, with the size and color of the rectangle related to the amplitude of the detected signal.  The beam enters the detector from the left.  The $\mu^-$ is seen as the long, forward-directed track.  In the side view (bottom), two photons from a $\pi^0$ decay are visible.  In the overhead view (top), one of the decay photons is unseen as its highly transverse trajectory stayed within a single $yz$ detector plane.}
\label{fig:TutEvd}
\end{figure*}

An illustrative simulated \thereaction{} resonant event as it would appear in the ND is shown in Fig.~\ref{fig:TutEvd}.  In this event, a $\pi^0$ decays promptly via $\pi^0\rightarrow\gamma\gamma$ (branching ratio 98.8$\%$ \cite{bib:PDG}).  Each decay photon produces an electromagnetic cascade.  In this example event, one of the photons is on a very transverse trajectory and is visible and
reconstructable only in the $yz$ view.

\subsection{Base Reconstruction}\label{sect:basereco}

A clustering algorithm~\cite{bib:DBSCAN} groups a collection of active detector cells (``hits'') nearby in space and time into a ``slice,'' intended to represent an individual neutrino interaction.  Within each slice, an interaction vertex is reconstructed by minimizing the angular spread of hits relative to the candidate vertex~\cite{bib:CHEP2015_reco}. Final state particles and electromagnetic cascades, as from photons, are reconstructed into ``prongs'' using a fuzzy $k$-means algorithm~\cite{bib:FuzzyKMeans} that clusters hits lying along a common direction relative to the vertex.  Separately, a Kalman filter algorithm~\cite{bib:Kalman} is applied to the event to better reconstruct muon-like tracks and to provide an energy estimate for them. A $k$-nearest-neighbors classifier from previous NOvA analyses~\cite{bib:RaddatzThesis} is used to identify the most muon-like track in each event, which is taken to be the $\mu^-$ candidate.

\subsection{Photon Identification}
\label{sect:photonreco}

\begin{figure}[tb!]
\centering
\hspace{-0.05in}\includegraphics[width=0.99\linewidth]{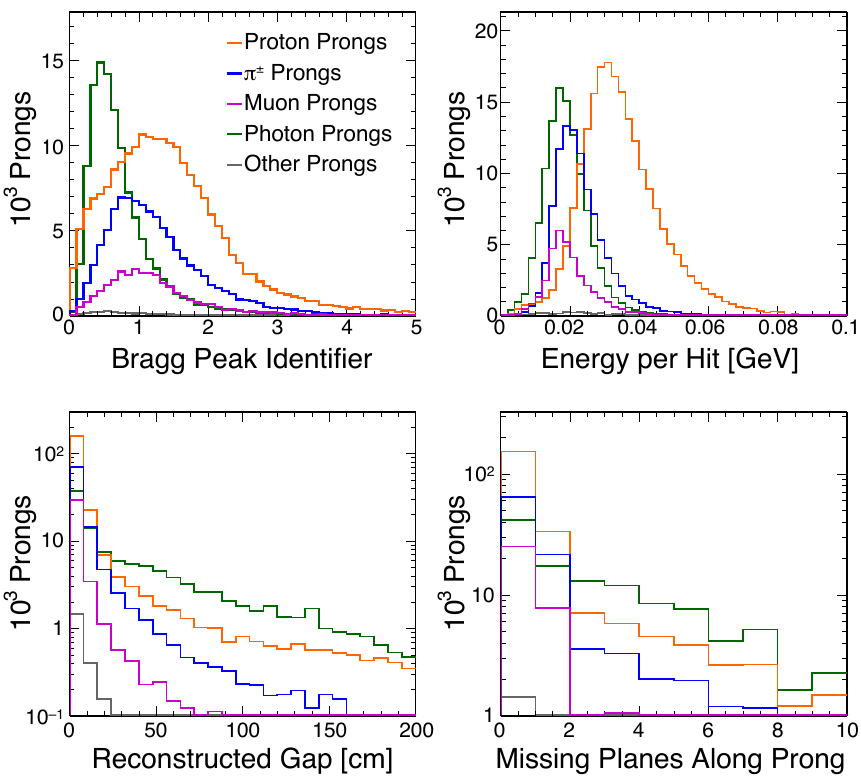}
\caption{Simulated distributions of the four variables used to identify photon-induced prongs.  ({\em n.b.}:\ The bottom two panels have logarithmic vertical scales.)}
\label{fig:SingProngVars}
\end{figure}

Candidate photon prongs are required to have a number of hits $N_\text{hit} \geq 10$ so that particle identification can be carried out effectively.  The calorimetric energy $E_\text{cal}$, defined as the sum of the calibrated energy deposited in all of the prong's hits, must also satisfy $E_\text{cal} > 100$\,MeV, roughly the energy deposited by a 10-hit minimum-ionizing track.   For prongs satisfying these thresholds, a likelihood ratio between photon and non-photon particle hypotheses is calculated based on the following four inputs:

\begin{enumerate}
\item Bragg peak identifier:\ ratio of average energy deposition in the furthest six hits from the prong start point (or five hits if $10\leq N_\text{hit}<12$) to the average energy deposition in the rest of the prong. This ratio gives a measure of the increase in $dE/dx$ towards the end of a prong.
\item Energy per hit:\ average calorimetric energy of all hits within the prong.
\item Reconstructed gap:\ distance from the reconstructed event vertex to the candidate prong's start point.
\item Missing planes along prong:\ largest number of consecutive planes without any energy deposition in the prong.
\end{enumerate}
Fig.~\ref{fig:SingProngVars} shows distributions of these four quantities for prongs in the simulated neutrino event sample. A number of familiar features can be seen.  Proton prongs exhibit the most prominent Bragg peak, while the electromagnetic cascades from photons score the lowest in that variable.  Protons have the highest $dE/dx$ (thus, highest energy per hit) of the listed particles.  Initial photons can lead to large prong-start gaps due to the radiation length in the detector, and the subsequent cascades can skip over planes due to secondary photons.

There are significant correlations between the Bragg peak identifier and the mean energy per hit and, separately, between the reconstructed gap and missing planes along the prong.  The correlations within each of these pairs are exploited when calculating the photon and non-photon likelihoods.  For numerical convenience, the logarithm of the likelihood ratio is used in calculations, equivalent to the difference in the logarithms of the likelihoods, $\Delta\log \mathcal{L}$.  An interaction-level score for events with at least two prongs is formed as
\begin{equation}
\text{CC}\pi^0\text{ID} = \max(\Delta \log{\mathcal{L}})\,,
\end{equation}
running over all prongs in the event that are not associated with the muon.  That is, the CC$\pi^0$ID score for the event is simply the highest photon-like score among all prongs, neglecting the identified muon prong.

Within an event, the prong that defines the \ccpi{}ID value is considered the photon candidate to be associated with the $\pi^0$ decay.  This photon candidate is used to reconstruct the $\pi^0$ kinematics.  No attempt is made to reconstruct a second shower from the decay as doing so gives increased background rates from prong combinatorics in events with significant hadronic activity and gives reduced signal efficiency at higher pion energies due to overlapping or energetically asymmetric photon pairs.  According to simulation, requiring a second prong would drop the signal purity to 29\% from the current 34\%.

\subsection{Reconstruction of Event Kinematics}

\begin{figure}[tb!]
\centering
\includegraphics[width=0.99\linewidth]{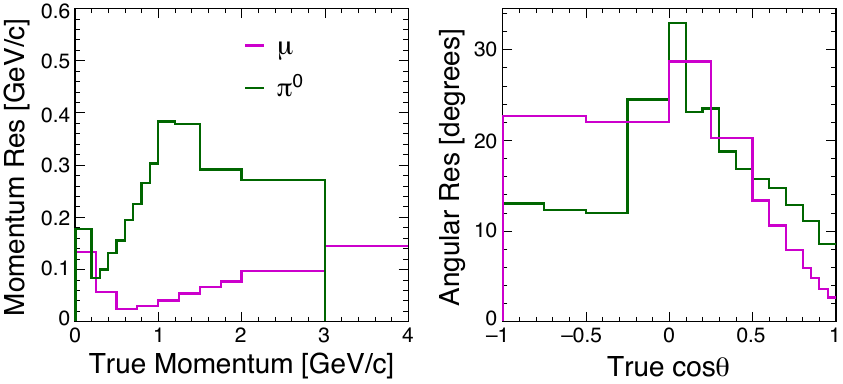}
\caption{The absolute momentum (left) and angular (right) resolutions for muons (magenta) and $\pi^0$'s (green) as predicted by the simulation.}
\label{fig:MuPiRes}
\end{figure}

The analysis utilizes reconstructed estimates of momentum and of angle with respect to the average beam direction for both the muon and the $\pi^0$.  The momentum of the $\pi^0$ is estimated as the $E_\text{cal}$ deposited by the single photon candidate prong.  This estimator is suitably close to the true $\pi^0$ momentum (into which it will ultimately be unfolded) given the predominance of overlapping (merged) or energetically asymmetric photon pairs~\cite{bib:PersheyThesis}.  The muon momentum is estimated as a linear function of the track length through the fully instrumented portion of the detector, $L_1$, and through the muon range stack, $L_2$, as
\begin{equation}
p_\mu=c_1L_1+c_2L_2\,.
\end{equation}
The constants $c_1$ and $c_2$ were determined by optimizing the simulated resolution~\footnote{All quoted resolutions are calculated as the RMS difference between reconstructed and true values in simulation.}, $3.5\%$ for muon momenta greater than 0.6\,$\mathrm{GeV}/c$.  The angular resolution is better than $10^\circ$ for $\cos\theta>0.5$. Muons with larger angles have poorer resolution due to their lower average energies. Those orthogonal to the beam direction ($\cos\theta\approx0$) are the most difficult to reconstruct due to the detector geometry.  The momentum and angular resolutions for $\pi^0$'s and muons are shown in Fig.~\ref{fig:MuPiRes}.

The neutrino energy is reconstructed as 
\begin{equation}
{E}_\nu = E_\mu + E_{\text{Had}}\,,
\end{equation}
where $E_\mu$ is reconstructed as described above.  The reconstructed hadronic energy, $E_{\text{Had}}$, is determined from the calorimetric energy of all hits except those in the muon track.  This estimate includes the $E_\text{cal}$ estimated from the $\pi^0$ candidate.  The neutrino energy resolution averaged over the sample is 9.5\%.

Differential cross sections in the kinematic variables $Q^2$ and $W$ are of interest.  These quantities are calculated via 
\begin{eqnarray}
Q^2 &=& -\left(P_\mu-P_\nu\right)^2\\
    &=&\frac{2E_\nu}{c}\left(\frac{E_\mu}{c}- p_\mu\cos\theta_{\mu}\right)-m_\mu^2 c^2
\label{eqn:Q2}
\end{eqnarray}
and
\begin{eqnarray}
W &=& \frac{1}{c}\lvert P_N+P_\nu-P_\mu\rvert \\
 &=& \frac{1}{c}\sqrt{m_N^2c^2-Q^2+2m_N(E_\nu-E_\mu)}\,,
\label{eqn:W}
\end{eqnarray}
where $m_N$ is the nucleon mass, taken numerically here to be the neutron mass, and where $P_i$ is the four-momentum of particle $i$.  Note that $p_i$ here represents the magnitude of the three-momentum of particle $i$, as it does throughout the text.  This expression assumes the struck target is a stationary nucleon which neglects Fermi motion within the nucleus and neutrino-parton scattering.  These kinematic variables as constructed are sensitive to underlying physics. The $W$ variable effectively distinguishes among $\Delta(1232)$ resonance, $N^\text{*}$ resonances, and deep inelastic scattering events.  These definitions of $Q^2$ and $W$, based on the true final state kinematics, are used in the cross-section definition rather than the event generator's values for these kinematic variables to reduce reliance on the generator's modeling of nuclear structure and FSI.

\subsection{Event Selection}\label{sec:selection}

To be analyzed, an interaction must produce a slice in the reconstruction. This requirement removes 2.7$\%$ of signal events.  Each slice is required to have at least 20 hits and to span at least four planes.  Events must also have a reconstructed vertex, track, and at least two reconstructed prongs.  The muon track and $\pi^0$ candidate prong are then chosen.

Reconstructed vertices are required to lie within a fiducial region defined as a 2\,m$\times$2\,m$\times$900\,cm box centered laterally in the detector and extending between 100\,cm and 1000\,cm from the front face.  This corresponds to a 35\,ton fiducial mass.  The fiducial region covers a relatively small fraction of the detector but ensures efficient containment of tracks.  The simulation predicts $2.8\times10^5$ signal events within the fiducial volume with a purity, defined as the fraction of the simulated sample that is signal, of $11.8\%$ with no further selection criteria applied.  There is a leakage of otherwise-selected fiducial \ccpi{} events out of the sample and a spillage of non-fiducial \ccpi{} events into the sample.  According to simulation, 2.6$\%$ of fiducial events leak out while 2.8$\%$ of the sample are non-fiducial events that spill in.  There are no kinematic differences between the leak-out and spill-in events.  As these two samples are similar, we treat the non-fiducial \ccpi{} events that leak into the selected sample as signal and thus as a direct compensation for the events that leak out.  The efficiency corrections applied in Sec.~\ref{sec:xsanalysis_eff} account for this.

Containment cuts are then applied to ensure reliable reconstruction of the muon and the photon.  The photon shower candidate's start and end points as well as the muon's start point are required to be well contained within the fully active portion of the detector.  The muon's end point is required to be well contained within any part of the detector including the muon range stack.  After containment cuts, the simulated efficiency is $32.8\%$ relative to all fiducial signal interactions, with a simulated purity of $22.6\%$.

\begin{figure}[tb!]
\centering
\includegraphics[width=0.99\linewidth]{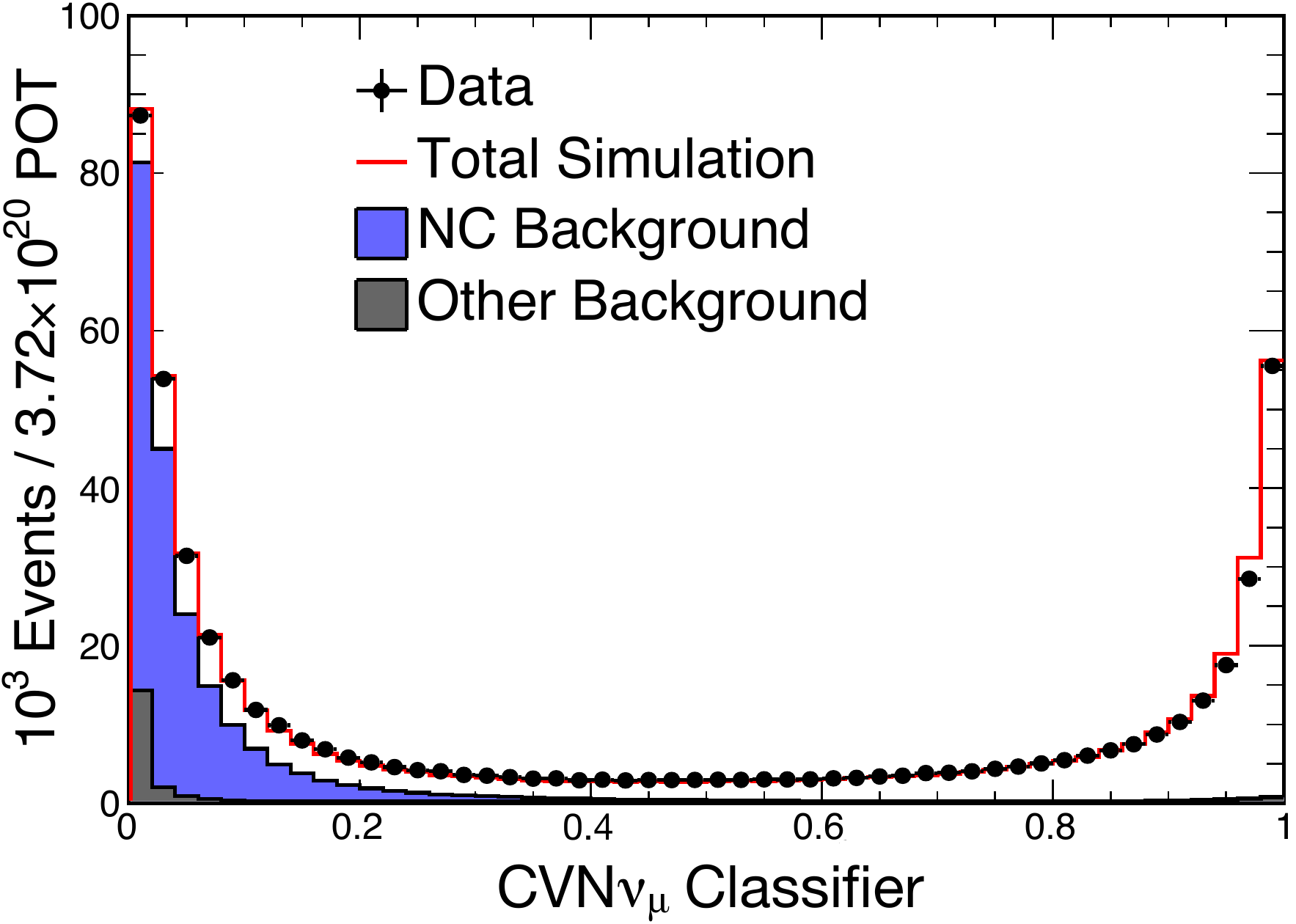}
\caption{Comparison of the data and simulation of the convolutional neural network-based CVN$\nu_\mu$ particle ID.  Backgrounds are indicated as shaded subsets of the total. The unshaded portion of the histogram represents $\nu_\mu$ CC events.  Note that the \ccpi{} signal is a subset of the unshaded portion of the histogram.}
\label{fig:CVN_DaMC}
\end{figure}

Events are then subject to a convolutional neural network identifier, CVN$\nu_\mu$ \cite{bib:CVN}.  This is a deep learning classifier used to separate $\nu_\mu$ CC events from the large NC background.  Events are required to have CVN$\nu_\mu$~$>0.5$, which leaves an NC contamination of 1.4$\%$ in the selected sample compared to 39.1$\%$ before the cut. 
The distributions of CVN$\nu_\mu$ for data and simulation for events at this stage of the selection are shown in Fig.~\ref{fig:CVN_DaMC}.

CVN$\nu_\mu$ also determines whether each event is most likely a quasi-elastic, resonant, deep inelastic scattering, or coherent interaction, trained on the GENIE labels for simulated events.  Events that CVN$\nu_\mu$ classifies as quasi-elastic or coherent are also rejected as these interactions have a well defined set of final-state particles and do not produce a $\pi^0$ except from FSI.  As a cross-check, this cut was replaced with a conventional quasi-elastic reduction cut by vetoing two-prong events whose measured momenta are consistent with the quasi-elastic formula.  The resulting cross sections agreed to within a percent, and the CVN$\nu_\mu$ cut is used as it more efficiently rejects background.

As noted previously, the photon candidate is required to have a calorimetric energy greater than 100 MeV and at least 10 hits to ensure reliable reconstruction.  Very loose preselection requirements are also applied on the input variables to \ccpiid{}.  These restrictions remove 2.9$\%$ of otherwise selected \ccpi{} events.

Lastly, events are removed if any reconstructed values for $E_\nu$, $p_\pi$, $p_\mu$, $Q^2$, and $W$ fail the signal definition conditions laid out in Table~\ref{table:KinRegion}.  This restriction removes 0.2$\%$ of otherwise selected \ccpi{} signal events.  

In total, the final selected sample consists of \ccpi{} signal (34.4\% of the total in the simulation); $\nu_\mu$ CC events without a $\pi^0$ emerging from the nucleus (62.1\%); non-$\nu_\mu$-CC events including NC events and CC events from other neutrino flavors present in the beam (3.2\%); and \ccpi{} events outside the kinematic limits of the signal definition (0.2\%). Note that the large background rate of $\nu_\mu$ CC events is expected in the sample since no explicit cut has been applied at this stage to reject such events.  Instead, the \ccpi{}ID event classifier is used to statistically separate signal and background in a fit to the classifier distribution, as described below.

\begin{figure}[tb!]
\centering
\includegraphics[width=0.99\linewidth]{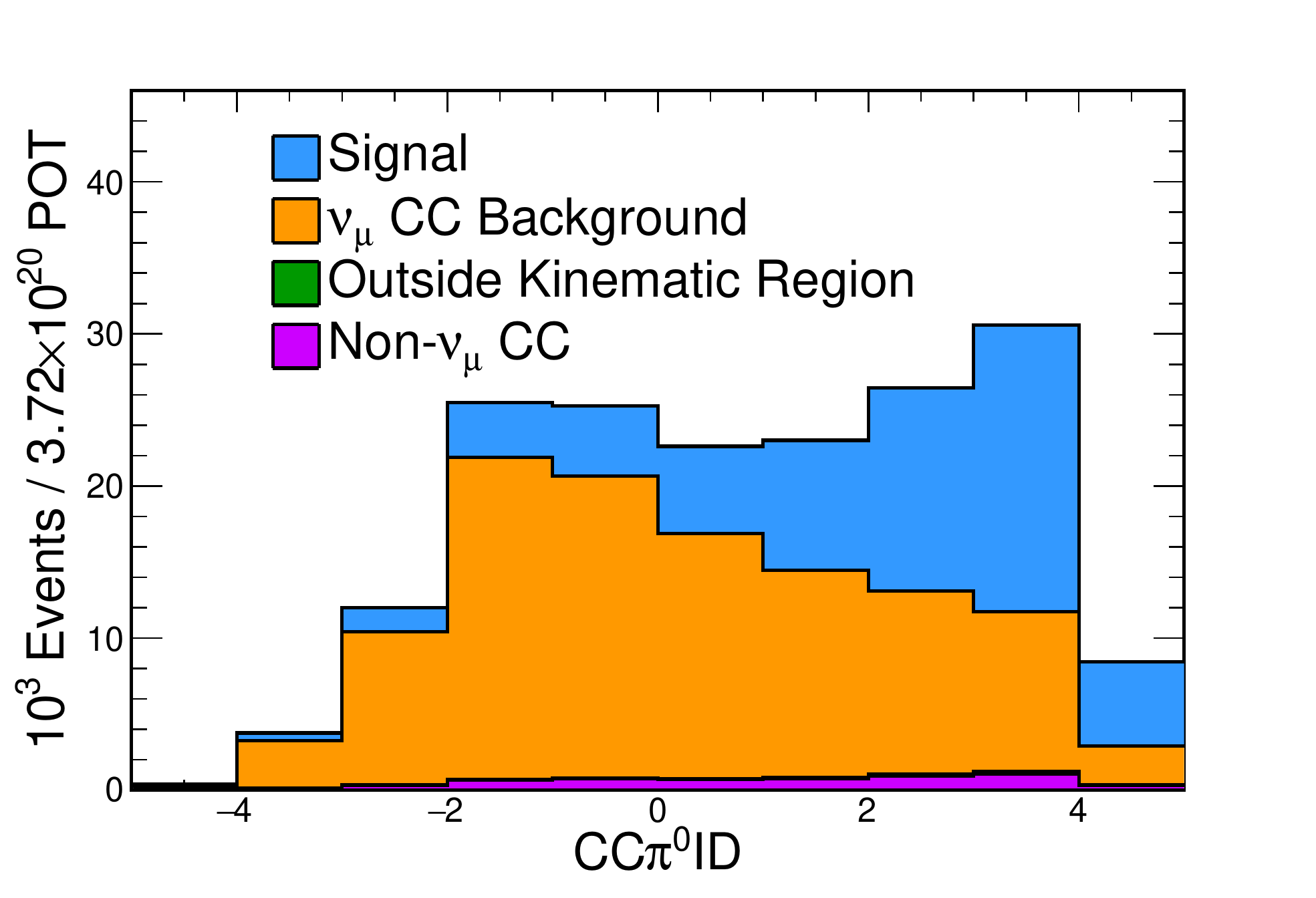}
\caption{The \ccpi{}ID distribution for all selected events in the simulation, showing clear discrimination between signal and backgrounds.  Backgrounds are primarily $\nu_\mu$~CC events, with small contributions from non-$\nu_\mu$~CC events and \ccpi{} events that fail the true kinematic restrictions. 
}
\label{fig:CCPi0ID}
\end{figure}

\begin{figure}[tb!]
\centering
\includegraphics[width=0.99\linewidth]{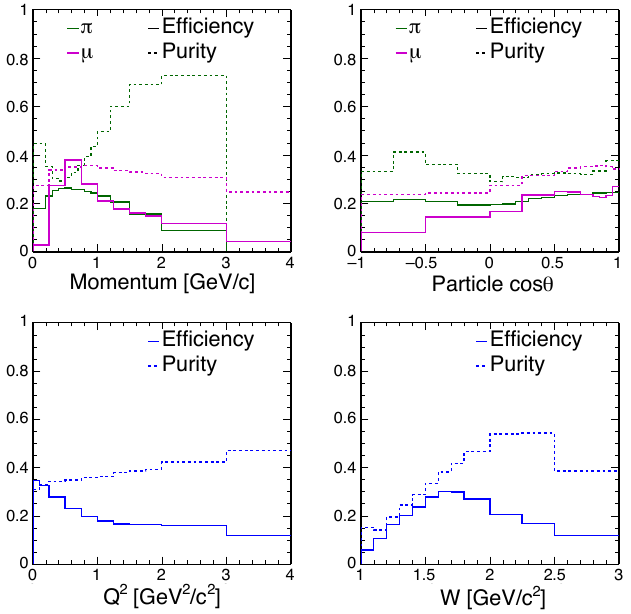}\caption{The event selection efficiency (solid) and purity (dashed) curves as a function of each measured variable.  The top plots show the particle momentum (left) and polar angle (right) with magenta and green curves describing the muon and $\pi^0$ kinematics, respectively.  The bottom plots give the efficiency and purity as a function of $Q^2$ (left) and $W$ (right).}
\label{fig:ObservableEffPur}
\end{figure}

The total selected event counts at each stage of the selection in data and simulation are shown in Table~\ref{table:AnalysisCutflow}.  At the data exposure of $3.72\times10^{20}$\,POT, there are 166,980 predicted and 164,871 observed events.  The simulated \ccpi{}ID distribution is plotted in Fig.~\ref{fig:CCPi0ID} after all selection cuts.  The overall signal efficiency is 21\% and the purity is 34\%, according to simulation.  The efficiency and purity as a function of the $\pi^0$ and $\mu^-$ kinematics are shown in Fig.~\ref{fig:ObservableEffPur}.  The efficiency drops sharply to 2.72$\%$ below 0.25\,$\mathrm{GeV}/c$ compared to 27.9$\%$ for $0.25<p_\mu<0.5\,\mathrm{GeV}/c$ due to difficulty in reconstructing short tracks and in $\mu^-/\pi^\pm$ discrimination.  Thus, the differential cross section for $p_\mu<0.25\,\mathrm{GeV}/c$ is not reported.  But, since only 0.5$\%$ of selected events have $p_\mu<0.25\,\mathrm{GeV}/c$, these events are not removed or treated as background when presenting cross-section results in any other kinematic variable.

\begin{table}[tb]
\caption{The event counts in data and predicted by simulation at various stages of the selection.  All simulated numbers are scaled to the data exposure of $3.72\times10^{20}$\,POT.  Signal efficiencies are calculated relative to interactions within the true fiducial volume.}

\centering
\begin{tabular}{ @{\hskip1pt}c@{\hskip1pt} | @{\hskip1pt}c@{\hskip1pt} | @{\hskip1pt}c@{\hskip1pt} | @{\hskip1pt}c@{\hskip1pt} | @{\hskip1pt}c@{\hskip1pt} }
Cut & $N_\text{evts}$ & $N_\text{GENIE}$ & Efficiency & Purity \\
\hline
Basic Reco / Fiducial & 2.192$\times10^6$ & 1.44$\times10^6$ & 85.1$\%$ & 16.3$\%$ \\
Containment & 517,317 & 400,797 & 32.8$\%$ & 22.6$\%$ \\
NC Rejection & 213,376 & 197,433 & 22.2$\%$ & 31.0$\%$ \\
QE/Coh Rejection & 197,858 & 186,779 & 22.0$\%$ & 32.7$\%$ \\
Prong Quality & 188,158 & 175,105 & 21.3$\%$ & 33.6$\%$ \\
Kinematic Restriction & 164,871 & 166,980 & 21.1$\%$ & 34.4$\%$ \\
\end{tabular}

\label{table:AnalysisCutflow}
\end{table}

\section{Cross-Section Analysis}\label{sec:xsanalysis}
Flux-averaged cross sections differential in final state kinematic variables are presented below.  Cross sections are extracted in bins of the true final-state-based kinematic variables.  These are defined by the equation 
\begin{equation}
\left(\frac{d\sigma}{dx}\right)_i = \frac{\left(U(\hat{S})\right)_i}{\Phi\,N_{T}\,\epsilon_i\,\Delta x_i}.
\end{equation}
Here, $x$ is the kinematic variable of interest, $i$ is the bin index, $\Phi$ is the integrated flux through the detector, $N_T$ is the number of interaction targets, $\epsilon_i$ is the detection efficiency in the bin, and $\Delta x_i$ is the bin width. $\hat{S}$ is a histogram that gives the signal estimate, using simulation constrained by data, in reconstructed bins of $x$.  $U$ refers to an unfolding procedure (described in Sec.~\ref{sec:Unfold}) which corrects for smearing effects and any estimator bias in the kinematic reconstruction process.  

\subsection{Constraint on Simulated Signal}

\label{sect:AnaFitting}

\begin{figure}[tb!]
\centering
\includegraphics[width=\linewidth]{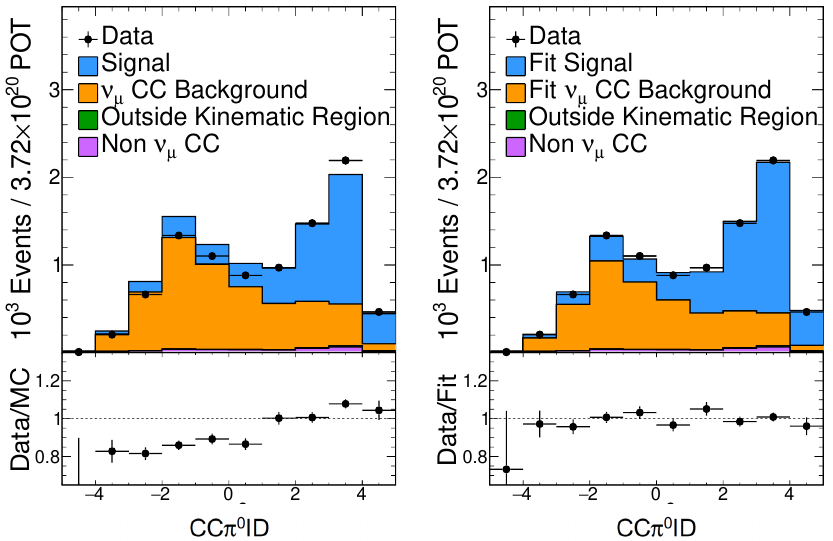}
\caption{An example fit to data in \ccpi{}ID for events reconstructed with 0.8 $< p_\pi <$ 1.0\,$\mathrm{GeV}/c$.  The left panel compares the unconstrained simulated \ccpi{}ID distribution and data while the right shows the simulation after constraining signal and background normalizations.}
\label{fig:CCPi0IDFitExamples}
\end{figure}

As shown in Fig.~\ref{fig:CCPi0ID}, the \ccpi{}ID distributions for the \ccpi{} signal and various categories of background each have their own distinctive shape. The signal and background normalizations are thus determined via a fit of the \ccpi{}ID distributions to the observed data.  The fit is performed in each kinematic bin independent of other bins. Of the three background categories, the non-$\nu_\mu$ CC background and the background of \ccpi{} that fail the kinematic requirements are both held fixed in the fit since they represent small populations, at 3.2$\%$ and 0.2$\%$, respectively.  The remaining ($\nu_\mu$ CC) background and signal normalizations are fit to the observed data distributions.  The signal and background normalizations float independently and without penalty.  As an example, the \ccpi{}ID distribution is shown in Fig.~\ref{fig:CCPi0IDFitExamples} for events in a representative bin of $p_\pi$ before and after fitting to data.

\subsection{Signal Unfolding}
\label{sec:Unfold}
For each variable of interest, an unfolding procedure is applied to correct for reconstruction effects and, thus, to obtain an estimate of each variable's true distribution.  The simulated true-to-reconstructed migration matrix for the variable is used as input to the unfolding.  Several unfolding procedures were considered, with a two-iteration D'Agostini~\cite{bib:DAgostini, bib:DAgostini2} technique ultimately being selected given its excellent robustness to the dominant analysis uncertainties, as demonstrated using sets of systematically fluctuated fake data.

\subsection{Nuclear Target Count}

\label{sect:Ana:TargetCount}

Since the NOvA detector consists of a mixture of materials, the result is presented as a cross section per nucleon.  The list of constituent elements is shown in Table~\ref{table:DetComposition}.  The detector is largely CH$_2$ with notable portions of oxygen, chlorine, and titanium, with $\langle A\rangle$~$=$~$15.96$.  The mass of the fiducial volume is 35,430 kg with a nucleon count of 2.12$\times10^{31}$.  This is known to better than 1$\%$ as described in Sec.~\ref{sect:normsyst}.

\begin{table}[tb]
\caption{The mass of the fiducial volume, broken down by element.  Trace amounts of nitrogen, sodium, sulphur, calcium, and tin are present and accounted for in simulation.}

\centering
\begin{tabular}{c | c | c | c }
Element & Mass [kg] & Nucleon Count 
& Mass Fraction\\
\hline
H & 3814.5 & $2.28\times10^{30}$ & 0.108 \\
C & 23650 & $1.41\times10^{31}$ & 0.667 \\
O & 1050 & $6.30\times10^{29}$ & 0.030 \\
Cl & 5690 & $3.40\times10^{30}$ & 0.161 \\
Ti & 1140 & $6.81\times10^{29}$ & 0.032\\
Other & 95 & $5.7\times10^{28}$ & 0.003
\end{tabular}

\label{table:DetComposition}
\end{table}

\subsection{Integrated Flux}

As described in Sec.~\ref{sec:BeamSim}, the NuMI flux simulation is constrained with external hadron production data using the PPFX package~\cite{bib:MINERvAPPFX}.  The integrated flux through the detector in the analyzed energy range is 87.0~$\nu_\mu$ / cm$^2$ / $10^{10}$\,POT.

\subsection{Efficiency Correction}
\label{sec:xsanalysis_eff}

Selection efficiencies for each of the variables are calculated with the simulation as a function of the true kinematics, shown in Fig.~\ref{fig:ObservableEffPur}.  The efficiency is defined as the ratio of selected signal events to the true number of signal events generated in the fiducial volume.  For the purposes of this calculation, the fiducial volume cut in the selection is truth based rather than reconstruction based to account for the compensation for those events that nominally leak out of the fiducial volume, as discussed in Sec.~\ref{sec:selection}.

\section{Systematic Uncertainty}

Several sources of systematic uncertainty are considered, classified into five separate groups: event normalization, neutrino flux, neutrino cross sections, uncertainty in the $\pi^\pm\rightarrow\pi^0$ CX cross section, and detector response.  For each source of uncertainty within these categories, cross-section covariance matrices are determined by repeating the cross-section measurement many times using an ensemble of altered versions of the simulation, where each alteration takes a random adjustment for the error source of interest chosen from its Gaussian distribution.

The final reported cross-section covariance matrices include these systematic uncertainties as well as (much smaller) statistical uncertainties.

\subsection{Data Normalization Uncertainties}
\label{sect:normsyst}

Three sources contribute to the normalization uncertainty. The fiducial mass, and thus the number of nucleon targets in the fiducial volume, is known to 0.7$\%$.  Variation in beam intensity, and thus also event pile-up, leads to no more than a $0.5\%$ effect on event reconstruction efficiency, as the event-isolating step of the reconstruction (Sec.~\ref{sect:basereco}) is highly effective.  A 1.9$\%$ uncertainty from modeling of particle containment is calculated by examining the differences of extracted cross sections between the inner and outer halves of the fiducial volume.  The uncertainty associated with POT counting and events interacting in the rock surrounding the detector were calculated but are negligible. The overall normalization uncertainty is 2.1$\%$.

\subsection{Flux Uncertainties}

Two broad sources of flux systematic uncertainty were assessed.  One comes from hadronization during the proton beam's initial collisions with the NuMI target.  The PPFX prediction \cite{bib:MINERvAPPFX} was used to calculate the corresponding uncertainty on the NuMI flux -- about $8\%$ near the beam peak.  Systematic uncertainties from beam transport were also assessed.  These correspond to multiple aspects of the neutrino beamline such as horn current, horn position, proton beam position on the target, beam spot size, and bending from the Earth's magnetic field in the decay pipe.  These transport effects taken together lead to flux uncertainties around 5\% near the beam peak.

\subsection{Neutrino Cross-Section Uncertainties}

The effect of neutrino interaction uncertainties is calculated using the GENIE event reweighting infrastructure~\cite{bib:GENIE2}.  Only systematic sources producing greater than a $0.5\%$ effect on the selected event rate are explicitly included in the analysis.  These include sources that affect GENIE's prediction for resonant and DIS events in the sample and modify the model by which initially produced particles undergo FSI.  The effect of any excluded GENIE systematic parameters was verified to have a negligible impact on the result.

Additionally, GENIE only calculates an uncertainty for DIS events with $W < 2\,\mathrm{GeV}/c^2$.  To treat the remaining DIS events, a 15$\%$ normalization uncertainty is added for DIS events at higher $W$, as motivated by neutrino scattering data~\cite{RevModPhys.70.1341}.

\subsection{Particle Tracking Cross-Section Uncertainties}

\begin{figure}[tb]
\centering
\hspace{-.1in}\includegraphics[width=0.95\linewidth]{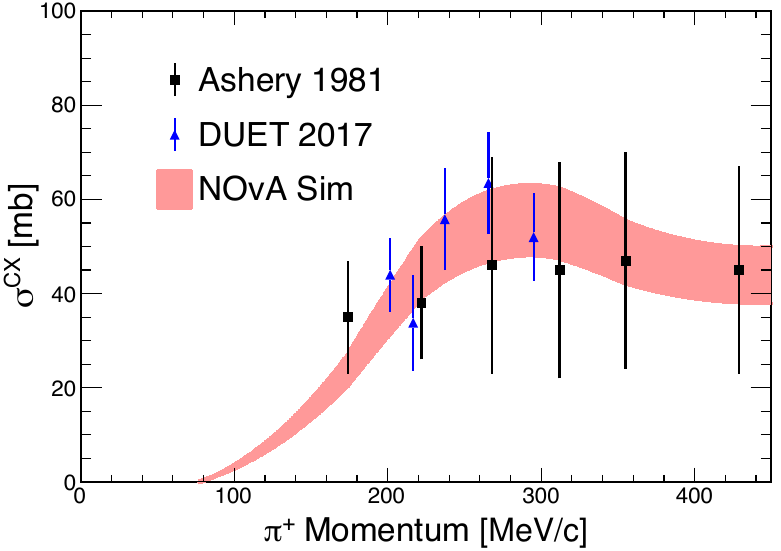}
\caption{A comparison of the Ashery \cite{bib:AsheryPion} and DUET \cite{bib:DUETPion} measurements of the $\pi^+\rightarrow\pi^0$ CX cross section and the spread in cross section of the NOvA simulation consistent with these results.  A 68.3$\%$ confidence interval for the range of scale factors applied to the nominal simulation was calculated as $1.061\pm0.146$ from a fit to the DUET data.}
\label{fig:SimEB_DUETCov_CEx}
\end{figure}

According to simulation, about a quarter of the $\nu_\mu$ CC background events ({\em cf.}\ Fig.~\ref{fig:CCPi0ID}) contain a secondary $\pi^0$ produced via hadronic interactions downstream in the detector, typically through the charge exchange (CX) reaction.  Given the importance of this process, a final systematic uncertainty adjusts the simulated cross section for $\pi^\pm\rightarrow\pi^0$ CX.  To bring in the most recent measurements of the CX cross section, the central value and error band used in this analysis were determined by fitting data from the DUET experiment \cite{bib:DUETPion}.  This data set offers a factor of three more precise cross section than that used to tune GEANT4 \cite{bib:AsheryPion}.  The fit increases the cross section by 6.1$\%$ relative to the default simulation with a 14.6$\%$ error band as shown in Fig.~\ref{fig:SimEB_DUETCov_CEx}.  The effect of uncertainty in the shape of this cross section was also studied but found to be negligible.

\subsection{Detector Response Uncertainties}
\label{syst:DetResponse}

Uncertainty in the light yield of the liquid scintillator is assessed by varying scintillator quenching parameters simulated according to the Birks-Chou formula \cite{bib:Birks, bib:BirksChou} given by
\begin{equation}
LY = A\frac{\frac{dE}{dx}}{1+k_B\frac{dE}{dx}+k_C\left(\frac{dE}{dx}\right)^2}\,.
\end{equation}
The Birks-Chou parameters were fit to data using the $dE/dX$ of stopping protons whose track length was greater than 1\,m.  This fit produced an anomalously high value $k_B$~$=$~$0.046$\,cm/MeV with $k_C$~$=$~$-0.00058$\,cm$^2$/MeV$^2$, versus typical values for organic liquid scintillators of $k_B \approx 0.01$\,cm/MeV.  Borexino measured $k_B=0.01055$\,cm/MeV \cite{bib:BorexinoBirks} in a similar scintillator blend to NOvA's.  The discrepancy is explained by the lack of a Cherenkov light component in the default simulation.  (Studies using a later version of the NOvA simulation that includes Cherenkov light yield $k_B$~$=$~$0.012$\,cm/MeV with $k_C=0$.)  To compensate for this known shortcoming of the simulation used here, a systematic uncertainty was calculated by taking the difference in the cross sections determined using the default simulation and a shifted simulation that assumes $k_B$~$=$~$0.012$\,cm/MeV.

\begin{figure}[tb!]
\includegraphics[width=0.45\textwidth]{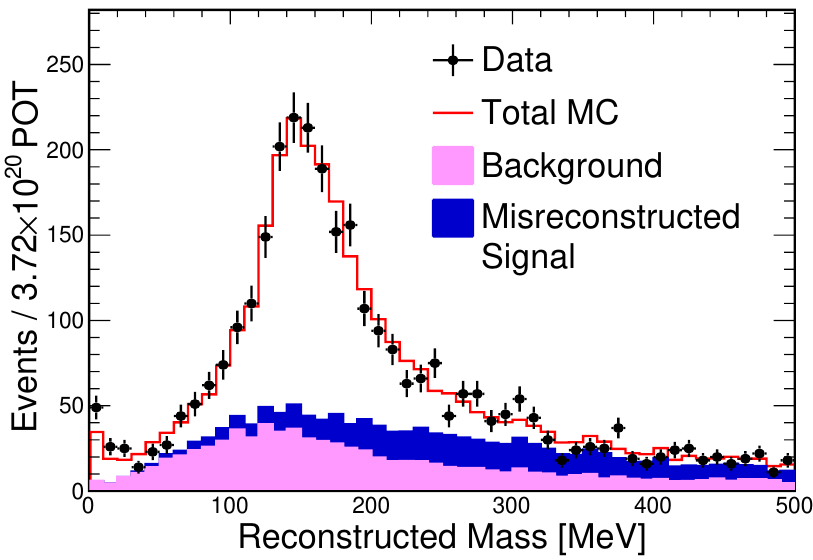}
\centering
\caption{The reconstructed $\pi^0$ mass from the small subsample of fully reconstructed $\mu\gamma\gamma$ candidate events.}
\label{fig:MassPeak}
\end{figure}

A set of detector calibration uncertainties are also included.  One class of uncertainties varies the energy of the photon candidate by 2$\%$ and the remaining hadronic energy by 5$\%$.  The hadronic energy uncertainty originates from a comparison of the simulated $dE/dx$ profile of stopping protons to data.  The photon energy uncertainty is determined by data/simulation comparisons of electrons from muon decay.  For photons, an additional gauge of the quality of the energy scale calibration is provided by the $\pi^0$ invariant mass, as reconstructed using a subset of the \ccpi{} sample with only three reconstructed particles consistent with a $\mu^-\gamma\gamma$ topology.  As shown in Fig.~\ref{fig:MassPeak}, the peak of the mass spectrum observed in data lies within 1$\%$ of the simulated peak, well within the assessed 2\% energy scale uncertainty.

A separate calibration uncertainty relates to the spatial variation in the collection and transport of light from energy depositions at various positions along the length of a detector cell.  While this uncertainty has a much smaller effect on the resulting cross-section measurement, its impact is not completely negligible in some regions of kinematic space, and thus it is included throughout.

\subsection{Tests of the Estimated Detector Response Error Band}
\label{sect:DetResponseTest}

High-purity samples of photons and protons were developed to test the estimated detector response systematic error band.  Photons are selected from a high-purity sample of two-prong NC$\pi^0$ events.  Both prongs are required to have $dE/dx$, reconstructed energy gaps, and prong length consistent with a photon.  Additionally, the reconstructed invariant mass must lie near $m_{\pi^0}$.  This selection gives an 82.1$\%$ pure sample of photons.  Protons are selected in two-prong events that pass a $\nu_\mu$ CC selection~\cite{bib:NOVA_SANumu}.  To identify events whose secondary prong was a proton, first, the reconstructed angle between the muon and proton candidate prongs must have $\cos\theta_{\mu p} > -0.8$ to remove a reconstruction failure that can split a muon track into two by mis-reconstructing the vertex.  Second, the proton direction inferred from the quasi-elastic formula \cite{bib:eVtoEeV} using only the observed muon kinematics is required to be coincident with the observed prong direction; the criterion is 
$\hat{\mathbf{p}}_p^{pr}\cdot\hat{\mathbf{p}}_p^{Q\!E}>0.9$, where $\hat{\mathbf{p}}_p^{pr}$ is the unit vector proton direction determined by the prong reconstruction and $\hat{\mathbf{p}}_p^{Q\!E}$ is that determined using the quasi-elastic formula.  This gives an 82.9$\%$ pure sample of protons.

\begin{figure}[tb!]
\centering
\includegraphics[width=0.99\linewidth]{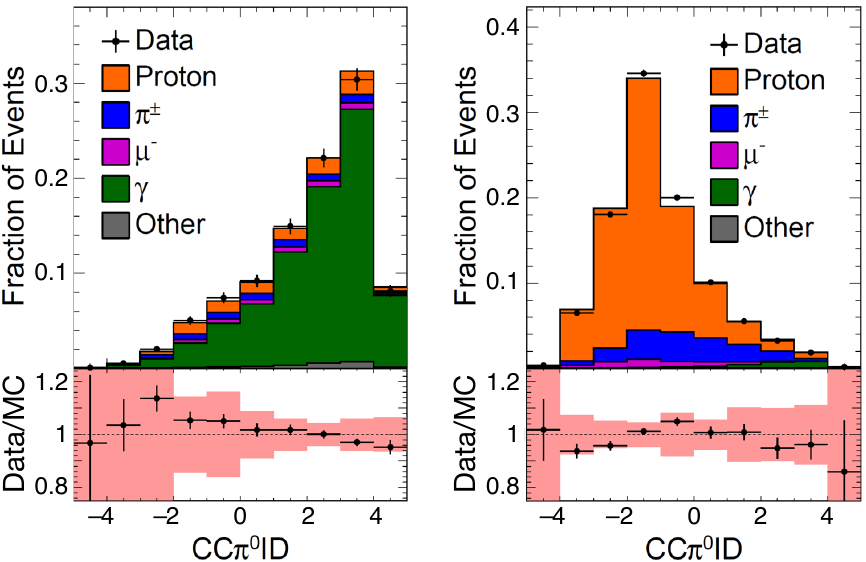}
\caption{The \ccpiid{} distribution for a control sample of photons (left) and protons (right).  Simulation and data have been area normalized to suppress overall normalization uncertainties.  Each bottom panel shows the ratio of data to simulation, with an error band from detector response uncertainties.}
\label{fig:Fig_DaMC_CCPi0ID_TestSamples}
\end{figure}

An area-normalized comparison of \ccpi{}ID in data and simulation with shape-only detector response systematic errors for these two samples is shown in Fig.~\ref{fig:Fig_DaMC_CCPi0ID_TestSamples}.  The differences observed between data and simulation lie well within the estimated error band, offering a level of confirmation that the systematic treatment adequately addresses the relevant uncertainties in the response of \ccpi{}ID to signal particles and the most important backgrounds.

\subsection{Total Systematic Uncertainty}

\begin{figure}[tb!]
\centering
\includegraphics[width=0.85\linewidth]{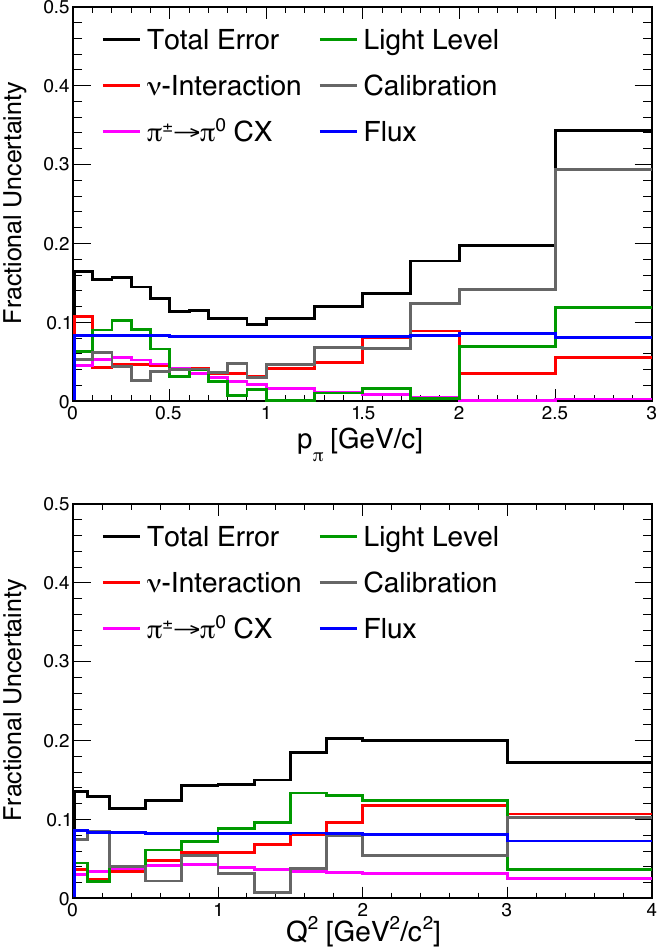}
\caption{The systematic uncertainty budget as a function of $\pi^0$ momentum (top) and $Q^2$ (bottom).  No single uncertainty source dominates the overall measurement uncertainty throughout the whole kinematic range.}
\label{fig:SystCocktail}
\end{figure}

The total systematic error, as a function of measured $p_\pi$ and $Q^2$, is shown in Fig.~\ref{fig:SystCocktail}.  The systematic uncertainty on the total cross section, broken down by each source, is shown in Table~\ref{table:TotSystEffects}.

\begin{table}[tb!]
\caption{The effect of each systematic uncertainty on the extracted total cross section.  The flux uncertainties are the largest source of systematic error, with large contributions from the light level, calibration, $\pi^\pm$ charge exchange, and the neutrino interaction model.}

\centering
\begin{tabular}{c | c }
Systematic Source & Rel. Error \\
\hline
Normalization & 2.1$\%$ \\
Flux & 8.3$\%$ \\
Neutrino Interaction Model & 4.6$\%$ \\
$\pi^\pm$ Charge Exchange  & 3.8$\%$ \\
Light Level & 6.8$\%$ \\
Calibration & 2.6$\%$ \\
\hline
Quadrature Sum & 12.5$\%$
\end{tabular}

\label{table:TotSystEffects}
\end{table}

\section{Analysis Results}

The sections that follow discuss the measured differential cross sections in each kinematic variable and the total cross section, with comparisons to the reference GENIE model throughout.  In general, a 7.5$\%$ larger total cross section is observed compared to the GENIE prediction, though results are within the systematic error associated with flux normalization.

\subsection{Muon Kinematics}

\begin{figure}[tb!]
\centering
\includegraphics[width=0.92\linewidth]{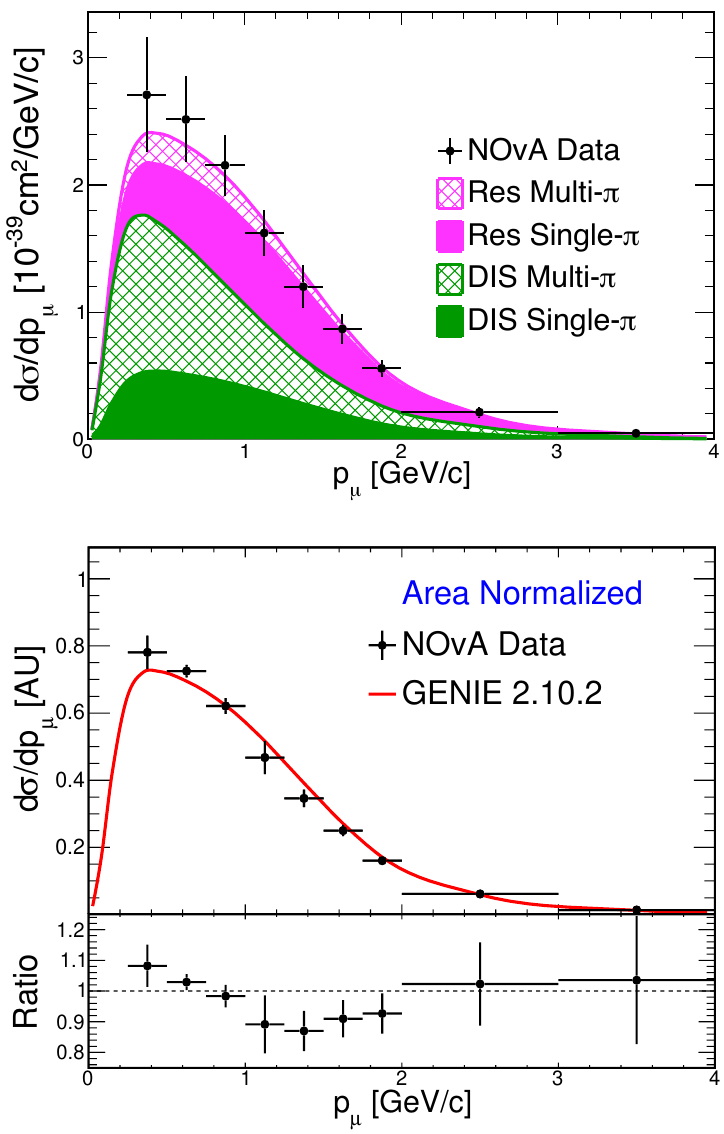}
\caption{The measured absolute differential cross section (top) and area-normalized differential cross section (bottom), per nucleon, vs.\ $p_\mu$. For the absolute cross section, the GENIE prediction is shown separated into resonant and DIS production and by $\pi$ multiplicity.  In the bottom panel, the ratio of the measured cross section to the GENIE prediction is also shown in a subpanel.}
\label{fig:xsec_pmu}
\end{figure}

\begin{figure}[tb!]
\centering
\includegraphics[width=0.92\linewidth]{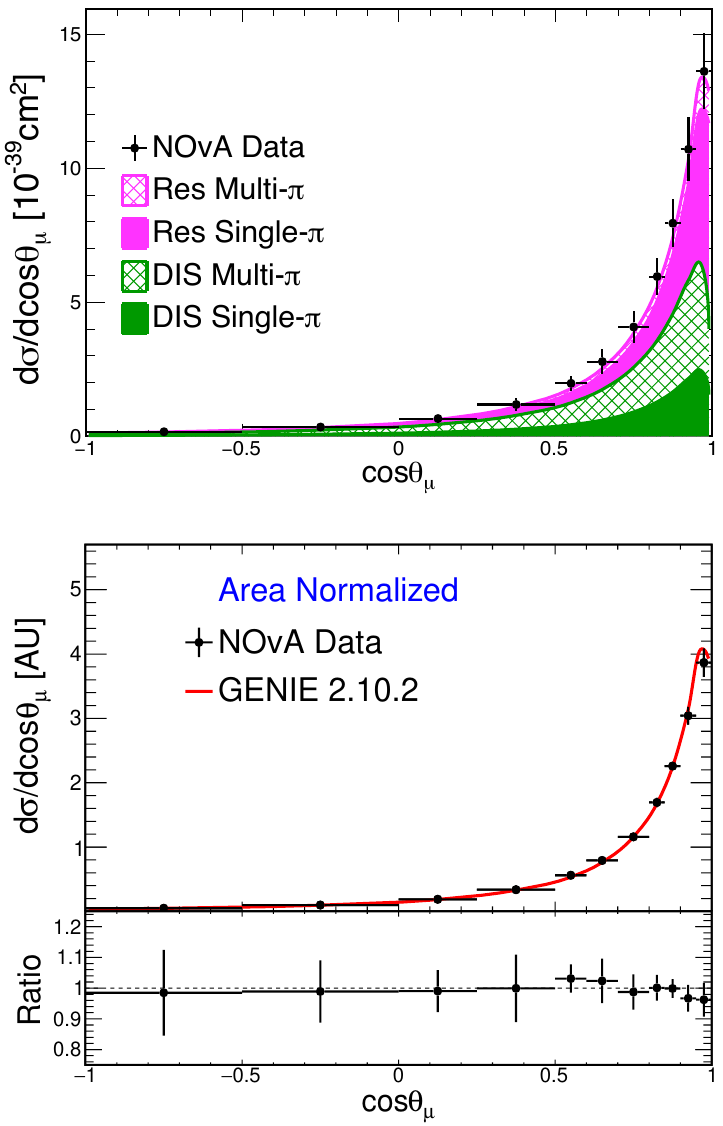}
\caption{As in Fig.~\ref{fig:xsec_pmu}, but for $\cos\theta_\mu$.}
\label{fig:xsec_cosmu}
\end{figure}

The measured differential cross sections in $p_\mu$ and $\cos\theta_\mu$, along with the GENIE predictions, are shown in Figs.~\ref{fig:xsec_pmu} and \ref{fig:xsec_cosmu}, respectively.  The predictions are separated into contributions from resonant and DIS scattering along with pion multiplicity.  Averaged over the flux,  multi-$\pi$ interactions account for 48$\%$ of the predicted total cross section.  Multi-$\pi$ events are more dominant at low $p_\mu$ where $\nu_\mu$~CC background events are more likely to be selected in $\nu_e$ oscillation measurements.  The cross section for $p_\mu<0.25\,\mathrm{GeV}/c$ is not reported due to the low efficiency for reconstructing and tagging short muon tracks.  GENIE predicts that this region represents 4$\%$ of the total cross section and is primarily populated by DIS multi-$\pi$ interactions.

In the lower panels the simulation is rescaled so that the integrated cross section matches the measurement, which allows trends in the comparison to be more readily observed.

A $\chi^2$ can be calculated for this result to explore the level of agreement between the measured and GENIE-predicted cross sections:
\begin{equation}
    \chi^2 = \sum d_iM_{ij}^{-1}d_j,
\end{equation}
where $d_i$ is the difference between measured and predicted cross sections in bin $i$ and $M_{ij}^{-1}$ is an element of the inverse of the covariance matrix.  The sum runs over all bins in the kinematic variable of interest.  The area scaling applied for visualization purposes above is not used here.  We calculate $\chi^2/{\rm\it dof}=9.75/9$ for the differential cross section in $p_\mu$ and $\chi^2/{\rm\it dof}=5.26/11$ for the differential cross section in $\cos\theta_\mu$.

\subsection{$\pi^0$ Kinematics}

\begin{figure}[!bt]
\centering
\includegraphics[width=0.92\linewidth]{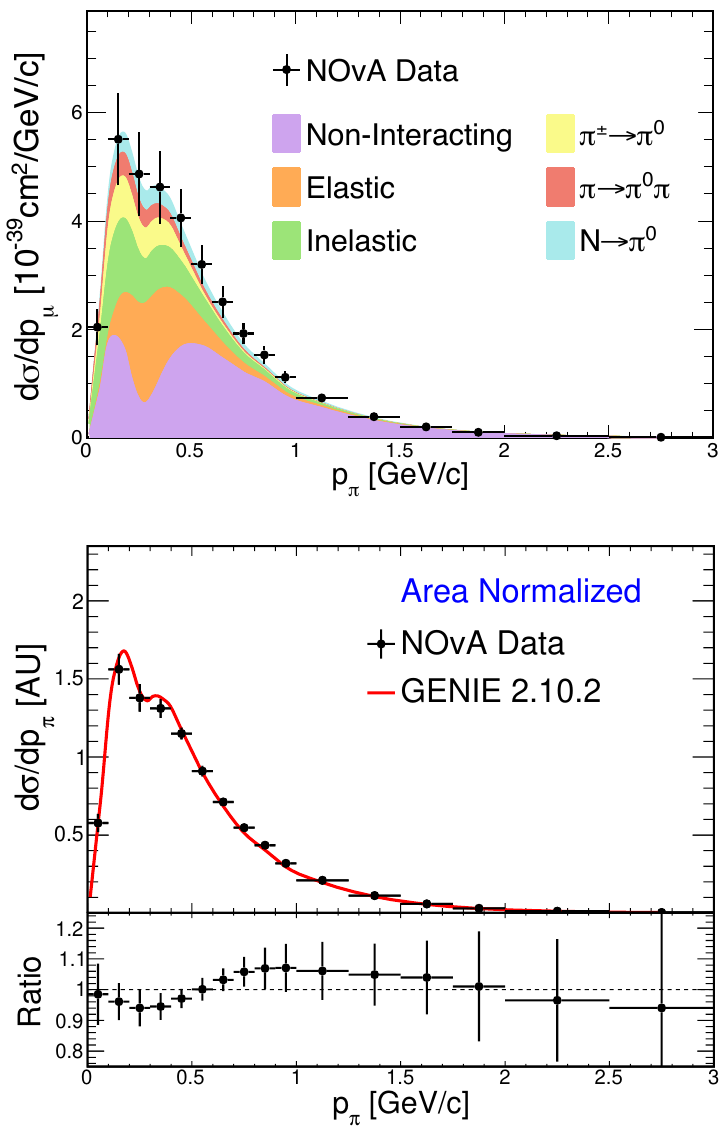}
\caption{The measured absolute differential cross section (top) and area-normalized differential cross section (bottom), per nucleon, vs.\ $p_\pi$. For the absolute cross section, the GENIE prediction is shown separated according to FSI channel; the $\pi+p\rightarrow\Delta(1232)$ resonance occurs around $p_\pi = 0.3\,\mathrm{GeV}/c$.  In the bottom panel, the ratio of the measured cross section to the GENIE prediction is also shown in a subpanel.}
\label{fig:xsec_ppi}
\end{figure}

\begin{figure}[tb!]
\centering
\includegraphics[width=0.92\linewidth]{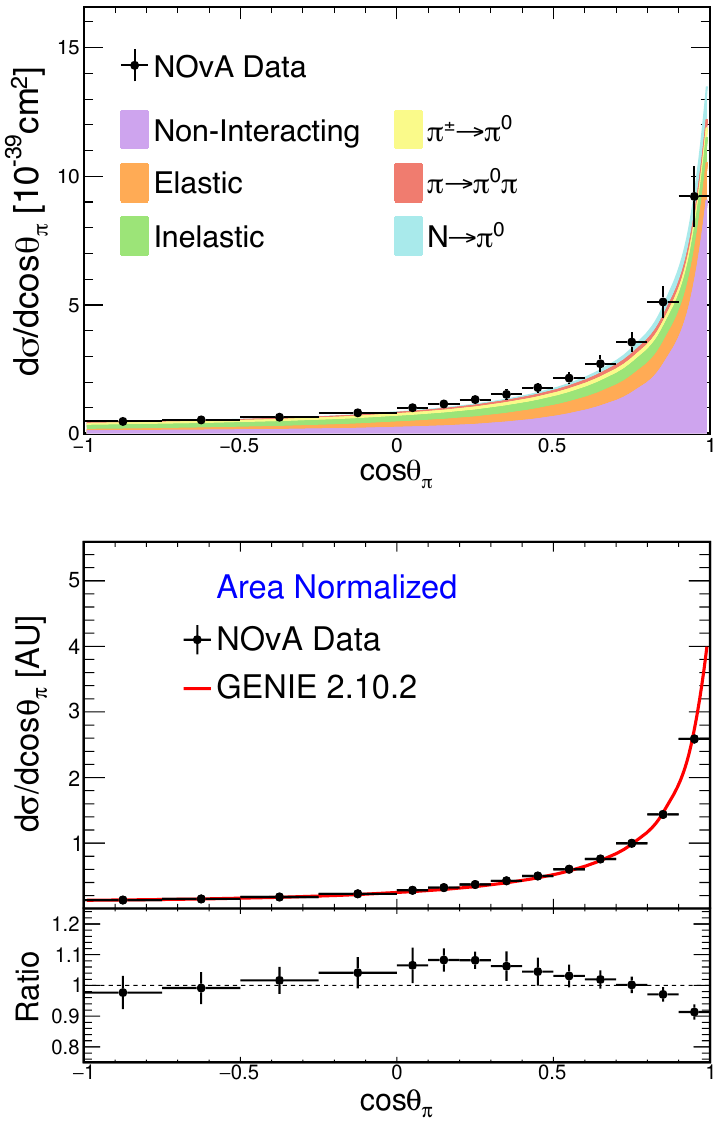}
\hskip 3.in
\caption{As in Fig.~\ref{fig:xsec_ppi}, but for $\cos\theta_\pi$.}
\label{fig:xsec_cospi}
\end{figure}

The differential cross sections in the $\pi^0$ kinematic variables are shown in Figs.~\ref{fig:xsec_ppi} and~\ref{fig:xsec_cospi}, with predictions separated into final-state interaction channels.  In the simulation, the majority of pions below $\sim$0.5~GeV/$c$ are involved in some sort of final-state interaction, and a subset of these involve production of a $\pi^0$ (namely the latter three FSI categories shown in the figures, corresponding to pion charge exchange, pion-induced $\pi^0$ production, and nucleon-induced $\pi^0$ production.)

The $p_\pi$ comparison yields a $\chi^2/{\rm\it dof}$~$=$~$21.49/16$.  There is a slight preference for a higher-momentum distribution in data, though consistent with the simulation given the uncertainty.  In the predicted $p_\pi$ differential cross section there is a clear dip near $p_\pi = 0.3\,\mathrm{GeV}/c$.  This stems primarily from  $\pi+p\rightarrow\Delta(1232)$ resonance production, which is modeled in GENIE alongside a number of other resonances and intranuclear hadronic processes.

For $\cos\theta_\pi$, a $\chi^2/{\rm\it dof}=32.12/14$ is calculated, with tension both in the $0$~$<$~$\cos\theta_\pi$~$<$~$0.5$ region and in the very forward-going direction.  Such a flattening of the peak could be evidence for stronger FSI than predicted by GENIE, though angular differences between DIS and resonant scattering, for instance, also influence this region.

\subsection{$Q^2$ and $W$} 

\begin{figure}[tb!]
\centering
\includegraphics[width=0.92\linewidth]{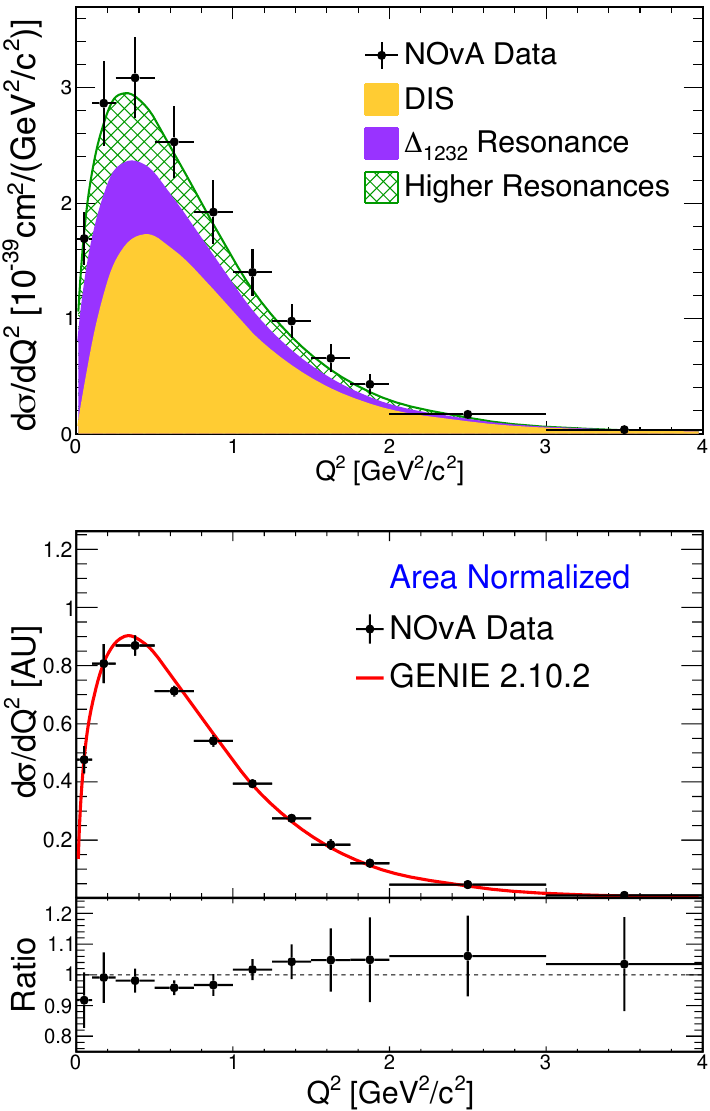}
\caption{The measured absolute differential cross section (top) and area-normalized differential cross section (bottom), per nucleon, vs.\ $Q^2$. For the absolute cross section, the GENIE prediction is shown separated into $\Delta(1232)$ resonance, $N^*$ resonances, and DIS contributions.  In the bottom panel, the ratio of the measured cross section to the GENIE prediction is also shown in a subpanel.}
\label{fig:xsec_q2}
\end{figure}

Results in $Q^2$, shown in Fig.~\ref{fig:xsec_q2}, agree
well with predictions with $\chi^2/dof$~$=$~$11.33/11$.  The prediction has been divided into contributions from DIS (60$\%$ of total cross section), the $\Delta(1232)$ resonance (22$\%$), and higher $N^*$ resonances (18$\%$).  This variable has shown sharp disagreements in past results \cite{bib:MINERvACCPi0-2017} when looking at single-$\pi$ events with very forward $\cos\theta_\mu$.

\begin{figure}[tb!]
\centering
\includegraphics[width=0.92\linewidth]{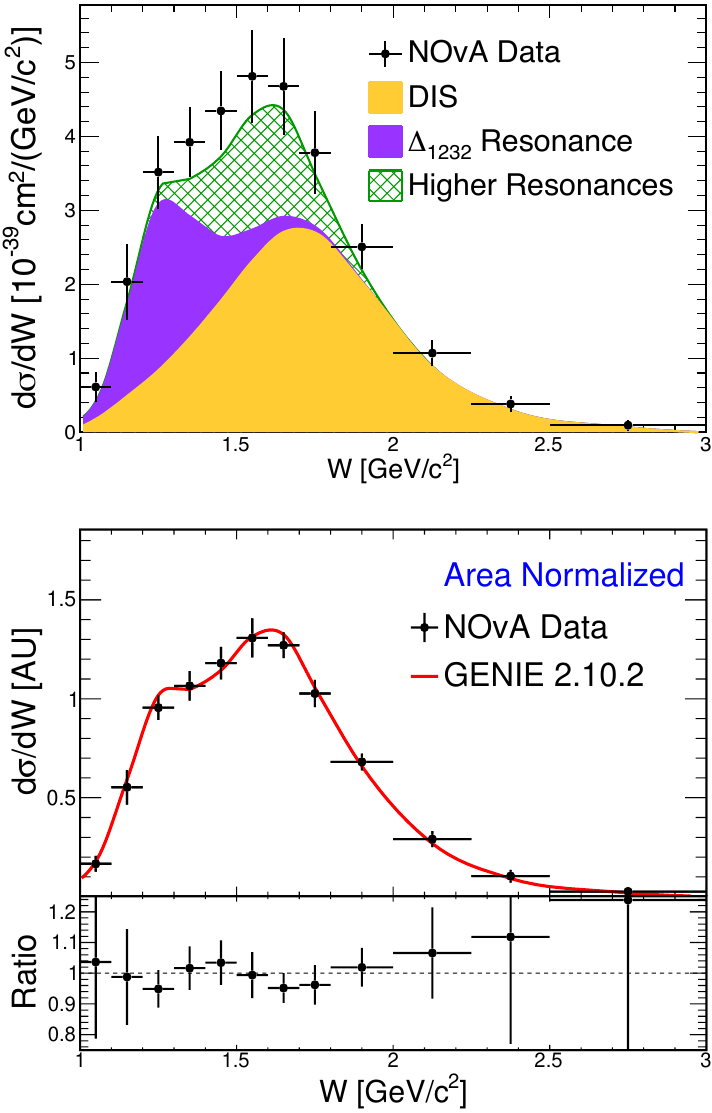}
\caption{As in Fig.~\ref{fig:xsec_q2}, but for $W$.}
\label{fig:xsec_W}
\end{figure}

The shape of the $W$ distribution (Fig.~\ref{fig:xsec_W}) is relatively well modeled, with $\chi^2/{\rm\it dof}=13.29/12$, particularly compared to other available $W$ measurements in semi-inclusive meson production measurements in other energy ranges \cite{bib:MiniBooNECCPi0, bib:MINERvACCPi0-2017}.  Notably, the observed shape is in agreement with GENIE for masses between 1.3 and 1.7\,GeV$/c^2$, exactly the region where GENIE predicts that $N^*$ resonances more massive than $\Delta(1232)$ contribute significantly to the cross section. This version of GENIE does not include interference effects between the various pion-production channels~\cite{PhysRevD.97.013002}.

\subsection{Total Flux-Averaged \ccpi{} Cross Section}

The total cross section determined by integrating the differential cross section in each kinematic variable is slightly different for each variable.  The reported total cross section is determined by averaging the total cross section obtained from the individual differential cross sections.  The differential cross section in $p_\mu$ is not included in the average as it is reported only for $p_\mu>0.25\,\mathrm{GeV}/c$.  The spread in individually measured cross sections is much smaller than the total cross section error, as shown in Table~\ref{table:TotIntSpread}.  The average total cross section is ($3.57\pm0.44)\times10^{-39}$\,cm$^2$ per nucleon.  

\begin{table}[tb!]
\caption{The total cross section and error as determined from each differential result.}

\centering
\begin{tabular}{ c | c  }
Kinematic Variable & $\langle\sigma\rangle_\Phi\, [10^{-39}\,\text{cm}^2]$ \\
\hline
$p_\pi$ & 3.53 $\pm$ 0.42 \\
$\cos\theta_\pi$ & 3.57 $\pm$ 0.42 \\
$\cos\theta_\mu$ & 3.52 $\pm$ 0.43 \\
$Q^2$ & 3.55 $\pm$ 0.44 \\
$W$ & 3.68 $\pm$ 0.43 \\
\hline
GENIE & 3.32 \\
\hline
\end{tabular}

\label{table:TotIntSpread}
\end{table}

\section{Conclusion}

A set of systematically limited measurements of $\pi^0$ production kinematics in $\nu_\mu$ CC events has been presented.  The measured total cross section is 7.5\% higher than the GENIE prediction but consistent within experimental error.  The studied energy region directly overlaps the transitional energy range between QE- and DIS-dominated scattering regimes so that baryon resonance and DIS events both contribute to the studied signal.  This energy region is particularly relevant for current and future oscillation measurements.  The signal definition for the measurement includes multi-$\pi$ events, which have been shown to cause the majority of $\pi^0$ background events in $\nu_\mu\rightarrow\nu_e$ oscillation measurements in NOvA.

Detailed numerical tables of the NOvA flux and the extracted cross sections with covariances are included as appendices.

\section{Acknowledgements}
This document was prepared by the NOvA collaboration using the resources of the Fermi National Accelerator Laboratory (Fermilab), a U.S. Department of Energy, Office of Science, HEP User Facility. Fermilab is managed by Fermi Research Alliance, LLC (FRA), acting under Contract No. DE-AC02-07CH11359. This work was supported by the U.S. Department of Energy; the U.S. National Science Foundation; the Department of Science and Technology, India; the European Research Council; the MSMT CR, GA UK, Czech Republic; the RAS, MSHE, and RFBR, Russia; CNPq and FAPEG, Brazil; UKRI, STFC and the Royal Society, United Kingdom; and the state and University of Minnesota.  We are grateful for the contributions of the staffs of the University of Minnesota at the Ash River Laboratory, and of Fermilab. For the purpose of open access, the author has applied a Creative Commons Attribution (CC BY) license to any Author Accepted Manuscript version arising.

\FloatBarrier

\bibliography{bibs}

\begin{thebibliography}{46}%
\makeatletter
\providecommand \@ifxundefined [1]{%
 \@ifx{#1\undefined}
}%
\providecommand \@ifnum [1]{%
 \ifnum #1\expandafter \@firstoftwo
 \else \expandafter \@secondoftwo
 \fi
}%
\providecommand \@ifx [1]{%
 \ifx #1\expandafter \@firstoftwo
 \else \expandafter \@secondoftwo
 \fi
}%
\providecommand \natexlab [1]{#1}%
\providecommand \enquote  [1]{``#1''}%
\providecommand \bibnamefont  [1]{#1}%
\providecommand \bibfnamefont [1]{#1}%
\providecommand \citenamefont [1]{#1}%
\providecommand \href@noop [0]{\@secondoftwo}%
\providecommand \href [0]{\begingroup \@sanitize@url \@href}%
\providecommand \@href[1]{\@@startlink{#1}\@@href}%
\providecommand \@@href[1]{\endgroup#1\@@endlink}%
\providecommand \@sanitize@url [0]{\catcode `\\12\catcode `\$12\catcode
  `\&12\catcode `\#12\catcode `\^12\catcode `\_12\catcode `\%12\relax}%
\providecommand \@@startlink[1]{}%
\providecommand \@@endlink[0]{}%
\providecommand \url  [0]{\begingroup\@sanitize@url \@url }%
\providecommand \@url [1]{\endgroup\@href {#1}{\urlprefix }}%
\providecommand \urlprefix  [0]{URL }%
\providecommand \Eprint [0]{\href }%
\providecommand \doibase [0]{http://dx.doi.org/}%
\providecommand \selectlanguage [0]{\@gobble}%
\providecommand \bibinfo  [0]{\@secondoftwo}%
\providecommand \bibfield  [0]{\@secondoftwo}%
\providecommand \translation [1]{[#1]}%
\providecommand \BibitemOpen [0]{}%
\providecommand \bibitemStop [0]{}%
\providecommand \bibitemNoStop [0]{.\EOS\space}%
\providecommand \EOS [0]{\spacefactor3000\relax}%
\providecommand \BibitemShut  [1]{\csname bibitem#1\endcsname}%
\let\auto@bib@innerbib\@empty
\bibitem [{\citenamefont {Acero}\ \emph {et~al.}(2018)\citenamefont {Acero}
  \emph {et~al.}}]{bib:NOvAJointFit}%
  \BibitemOpen
  \bibfield  {author} {\bibinfo {author} {\bibfnamefont {M.~A.}\ \bibnamefont
  {Acero}} \emph {et~al.} (\bibinfo {collaboration} {NOvA Collaboration}),\
  }\href {\doibase 10.1103/PhysRevD.98.032012} {\bibfield  {journal} {\bibinfo
  {journal} {Phys. Rev. D}\ }\textbf {\bibinfo {volume} {98}},\ \bibinfo
  {pages} {032012} (\bibinfo {year} {2018})}\BibitemShut {NoStop}%
\bibitem [{\citenamefont {Abi}\ \emph {et~al.}(2020)\citenamefont {Abi} \emph
  {et~al.}}]{bib:DUNETDR}%
  \BibitemOpen
  \bibfield  {author} {\bibinfo {author} {\bibfnamefont {B.}~\bibnamefont
  {Abi}} \emph {et~al.} (\bibinfo {collaboration} {DUNE Collaboration}),\
  }\href@noop {} {\  (\bibinfo {year} {2020})},\ \Eprint
  {http://arxiv.org/abs/2002.03005} {arXiv:2002.03005 [hep-ex]} \BibitemShut
  {NoStop}%
\bibitem [{\citenamefont {Radecky}\ \emph {et~al.}(1982)\citenamefont {Radecky}
  \emph {et~al.}}]{bib:ANLCCPi0}%
  \BibitemOpen
  \bibfield  {author} {\bibinfo {author} {\bibfnamefont {G.~M.}\ \bibnamefont
  {Radecky}} \emph {et~al.} (\bibinfo {collaboration} {ANL}),\ }\href {\doibase
  10.1103/PhysRevD.25.1161} {\bibfield  {journal} {\bibinfo  {journal} {Phys.
  Rev. D}\ }\textbf {\bibinfo {volume} {25}},\ \bibinfo {pages} {1161}
  (\bibinfo {year} {1982})}\BibitemShut {NoStop}%
\bibitem [{\citenamefont {Kitagaki}\ \emph {et~al.}(1986)\citenamefont
  {Kitagaki} \emph {et~al.}}]{bib:BNLCCPi0}%
  \BibitemOpen
  \bibfield  {author} {\bibinfo {author} {\bibfnamefont {T.}~\bibnamefont
  {Kitagaki}} \emph {et~al.} (\bibinfo {collaboration} {BNL}),\ }\href
  {\doibase 10.1103/PhysRevD.34.2554} {\bibfield  {journal} {\bibinfo
  {journal} {Phys. Rev. D}\ }\textbf {\bibinfo {volume} {34}},\ \bibinfo
  {pages} {2554} (\bibinfo {year} {1986})}\BibitemShut {NoStop}%
\bibitem [{\citenamefont {Le}\ \emph {et~al.}(2015)\citenamefont {Le} \emph
  {et~al.}}]{bib:MINERvACCPi0-2015}%
  \BibitemOpen
  \bibfield  {author} {\bibinfo {author} {\bibfnamefont {T.}~\bibnamefont {Le}}
  \emph {et~al.} (\bibinfo {collaboration} {MINERvA Collaboration}),\ }\href
  {\doibase 10.1016/j.physletb.2015.07.039} {\bibfield  {journal} {\bibinfo
  {journal} {Phys. Lett. B}\ }\textbf {\bibinfo {volume} {749}},\ \bibinfo
  {pages} {130} (\bibinfo {year} {2015})}\BibitemShut {NoStop}%
\bibitem [{\citenamefont {Altinok}\ \emph {et~al.}(2017)\citenamefont {Altinok}
  \emph {et~al.}}]{bib:MINERvACCPi0-2017}%
  \BibitemOpen
  \bibfield  {author} {\bibinfo {author} {\bibfnamefont {O.}~\bibnamefont
  {Altinok}} \emph {et~al.} (\bibinfo {collaboration} {MINERvA
  Collaboration}),\ }\href {\doibase 10.1103/PhysRevD.96.072003} {\bibfield
  {journal} {\bibinfo  {journal} {Phys. Rev. D}\ }\textbf {\bibinfo {volume}
  {96}},\ \bibinfo {pages} {072003} (\bibinfo {year} {2017})}\BibitemShut
  {NoStop}%
\bibitem [{\citenamefont {Coplowe}\ \emph {et~al.}(2020)\citenamefont {Coplowe}
  \emph {et~al.}}]{PhysRevD.102.072007}%
  \BibitemOpen
  \bibfield  {author} {\bibinfo {author} {\bibfnamefont {D.}~\bibnamefont
  {Coplowe}} \emph {et~al.} (\bibinfo {collaboration}
  {$\mathrm{MINER}\ensuremath{\nu}\mathrm{A}$ Collaboration}),\ }\href
  {\doibase 10.1103/PhysRevD.102.072007} {\bibfield  {journal} {\bibinfo
  {journal} {Phys. Rev. D}\ }\textbf {\bibinfo {volume} {102}},\ \bibinfo
  {pages} {072007} (\bibinfo {year} {2020})}\BibitemShut {NoStop}%
\bibitem [{\citenamefont {Aguilar-Arevalo}\ \emph {et~al.}(2011)\citenamefont
  {Aguilar-Arevalo} \emph {et~al.}}]{bib:MiniBooNECCPi0}%
  \BibitemOpen
  \bibfield  {author} {\bibinfo {author} {\bibfnamefont {A.~A.}\ \bibnamefont
  {Aguilar-Arevalo}} \emph {et~al.} (\bibinfo {collaboration} {MiniBooNE
  Collaboration}),\ }\href {\doibase 10.1103/PhysRevD.83.052009} {\bibfield
  {journal} {\bibinfo  {journal} {Phys. Rev. D}\ }\textbf {\bibinfo {volume}
  {83}},\ \bibinfo {pages} {052009} (\bibinfo {year} {2011})}\BibitemShut
  {NoStop}%
\bibitem [{\citenamefont {Mariani}(2007)}]{bib:K2KCCPi0}%
  \BibitemOpen
  \bibfield  {author} {\bibinfo {author} {\bibfnamefont {C.}~\bibnamefont
  {Mariani}} (\bibinfo {collaboration} {K2K Collaboration}),\ }\href {\doibase
  10.1063/1.2834471} {\bibfield  {journal} {\bibinfo  {journal} {AIP Conf.
  Proc.}\ }\textbf {\bibinfo {volume} {967}},\ \bibinfo {pages} {174} (\bibinfo
  {year} {2007})}\BibitemShut {NoStop}%
\bibitem [{\citenamefont {Adams}\ \emph {et~al.}(2019)\citenamefont {Adams}
  \emph {et~al.}}]{PhysRevD.99.091102}%
  \BibitemOpen
  \bibfield  {author} {\bibinfo {author} {\bibfnamefont {C.}~\bibnamefont
  {Adams}} \emph {et~al.} (\bibinfo {collaboration} {MicroBooNE
  Collaboration}),\ }\href {\doibase 10.1103/PhysRevD.99.091102} {\bibfield
  {journal} {\bibinfo  {journal} {Phys. Rev. D}\ }\textbf {\bibinfo {volume}
  {99}},\ \bibinfo {pages} {091102} (\bibinfo {year} {2019})}\BibitemShut
  {NoStop}%
\bibitem [{\citenamefont {Adamson}\ \emph {et~al.}(2016)\citenamefont {Adamson}
  \emph {et~al.}}]{bib:NuMI}%
  \BibitemOpen
  \bibfield  {author} {\bibinfo {author} {\bibfnamefont {P.}~\bibnamefont
  {Adamson}} \emph {et~al.},\ }\href {\doibase
  https://doi.org/10.1016/j.nima.2015.08.063} {\bibfield  {journal} {\bibinfo
  {journal} {Nucl. Instrum. Meth. A}\ }\textbf {\bibinfo {volume} {806}},\
  \bibinfo {pages} {279} (\bibinfo {year} {2016})}\BibitemShut {NoStop}%
\bibitem [{\citenamefont {Mufson}\ \emph {et~al.}(2015)\citenamefont {Mufson}
  \emph {et~al.}}]{bib:Scintillator}%
  \BibitemOpen
  \bibfield  {author} {\bibinfo {author} {\bibfnamefont {S.}~\bibnamefont
  {Mufson}} \emph {et~al.},\ }\href {\doibase
  https://doi.org/10.1016/j.nima.2015.07.026} {\bibfield  {journal} {\bibinfo
  {journal} {Nucl. Instrum. Meth. A}\ }\textbf {\bibinfo {volume} {799}},\
  \bibinfo {pages} {1 } (\bibinfo {year} {2015})}\BibitemShut {NoStop}%
\bibitem [{\citenamefont {B{\"o}hlen}\ \emph {et~al.}(2014)\citenamefont
  {B{\"o}hlen} \emph {et~al.}}]{bib:FLUKA2}%
  \BibitemOpen
  \bibfield  {author} {\bibinfo {author} {\bibfnamefont {T.}~\bibnamefont
  {B{\"o}hlen}} \emph {et~al.},\ }\href {\doibase
  https://doi.org/10.1016/j.nds.2014.07.049} {\bibfield  {journal} {\bibinfo
  {journal} {Nucl. Data Sheets}\ }\textbf {\bibinfo {volume} {120}},\ \bibinfo
  {pages} {211 } (\bibinfo {year} {2014})}\BibitemShut {NoStop}%
\bibitem [{\citenamefont {Ferrari}\ \emph {et~al.}(2005)\citenamefont {Ferrari}
  \emph {et~al.}}]{bib:FLUKA1}%
  \BibitemOpen
  \bibfield  {author} {\bibinfo {author} {\bibfnamefont {A.}~\bibnamefont
  {Ferrari}} \emph {et~al.},\ }\href@noop {} {}\bibinfo {type} {Tech. Rep.}\
  \bibinfo {number} {CERN-2005-010, SLAC-R-773, INFN-TC-05-11}\ (\bibinfo
  {year} {2005})\BibitemShut {NoStop}%
\bibitem [{\citenamefont {Campanella}\ \emph {et~al.}(1999)\citenamefont
  {Campanella} \emph {et~al.}}]{bib:FLUGG}%
  \BibitemOpen
  \bibfield  {author} {\bibinfo {author} {\bibfnamefont {M.}~\bibnamefont
  {Campanella}} \emph {et~al.},\ }\href@noop {} {}\bibinfo {type} {Tech. Rep.}\
  \bibinfo {number} {ATL-SOFT-99-004, ATL-COM-SOFT-99-004,
  CERN-ATL-SOFT-99-004}\ (\bibinfo {year} {1999})\BibitemShut {NoStop}%
\bibitem [{\citenamefont {Aliaga}\ \emph {et~al.}(2016)\citenamefont {Aliaga}
  \emph {et~al.}}]{bib:MINERvAPPFX}%
  \BibitemOpen
  \bibfield  {author} {\bibinfo {author} {\bibfnamefont {L.}~\bibnamefont
  {Aliaga}} \emph {et~al.} (\bibinfo {collaboration} {MINERvA Collaboration}),\
  }\href {\doibase 10.1103/PhysRevD.94.092005} {\bibfield  {journal} {\bibinfo
  {journal} {Phys. Rev. D}\ }\textbf {\bibinfo {volume} {94}},\ \bibinfo
  {pages} {092005} (\bibinfo {year} {2016})},\ \bibinfo {note} {[Addendum:
  Phys. Rev. D \textbf{95} (2017), no.3, 039903]}\BibitemShut {NoStop}%
\bibitem [{\citenamefont {Andreopoulos}\ \emph {et~al.}(2010)\citenamefont
  {Andreopoulos} \emph {et~al.}}]{bib:GENIE1}%
  \BibitemOpen
  \bibfield  {author} {\bibinfo {author} {\bibfnamefont {C.}~\bibnamefont
  {Andreopoulos}} \emph {et~al.} (\bibinfo {collaboration} {GENIE
  Collaboration}),\ }\href {\doibase 10.1016/j.nima.2009.12.009} {\bibfield
  {journal} {\bibinfo  {journal} {Nucl. Instrum. Meth. A}\ }\textbf {\bibinfo
  {volume} {614}},\ \bibinfo {pages} {87} (\bibinfo {year} {2010})}\BibitemShut
  {NoStop}%
\bibitem [{\citenamefont {Dytman}(2007)}]{Dytman:2007zz}%
  \BibitemOpen
  \bibfield  {author} {\bibinfo {author} {\bibfnamefont {S.}~\bibnamefont
  {Dytman}},\ }\href {\doibase 10.1063/1.2720468} {\bibfield  {journal}
  {\bibinfo  {journal} {AIP Conf. Proc.}\ }\textbf {\bibinfo {volume} {896}},\
  \bibinfo {pages} {178} (\bibinfo {year} {2007})}\BibitemShut {NoStop}%
\bibitem [{\citenamefont {Dytman}(2009)}]{Dytman:2009zz}%
  \BibitemOpen
  \bibfield  {author} {\bibinfo {author} {\bibfnamefont {S.}~\bibnamefont
  {Dytman}},\ }\href@noop {} {\bibfield  {journal} {\bibinfo  {journal} {Acta
  Phys. Polon. B}\ }\textbf {\bibinfo {volume} {40}},\ \bibinfo {pages} {2445}
  (\bibinfo {year} {2009})}\BibitemShut {NoStop}%
\bibitem [{\citenamefont {Llewellyn~Smith}(1972)}]{bib:LlewellynSmith}%
  \BibitemOpen
  \bibfield  {author} {\bibinfo {author} {\bibfnamefont {C.~H.}\ \bibnamefont
  {Llewellyn~Smith}},\ }\href {\doibase 10.1016/0370-1573(72)90010-5}
  {\bibfield  {journal} {\bibinfo  {journal} {Phys. Rept.}\ }\textbf {\bibinfo
  {volume} {3}},\ \bibinfo {pages} {261} (\bibinfo {year} {1972})}\BibitemShut
  {NoStop}%
\bibitem [{\citenamefont {Rein}\ and\ \citenamefont
  {Sehgal}(1981)}]{bib:ResReinSeghal}%
  \BibitemOpen
  \bibfield  {author} {\bibinfo {author} {\bibfnamefont {D.}~\bibnamefont
  {Rein}}\ and\ \bibinfo {author} {\bibfnamefont {L.~M.}\ \bibnamefont
  {Sehgal}},\ }\href {\doibase https://doi.org/10.1016/0003-4916(81)90242-6}
  {\bibfield  {journal} {\bibinfo  {journal} {Ann. Phys.}\ }\textbf {\bibinfo
  {volume} {133}},\ \bibinfo {pages} {79 } (\bibinfo {year}
  {1981})}\BibitemShut {NoStop}%
\bibitem [{\citenamefont {Bodek}\ and\ \citenamefont
  {Yang}(2003)}]{bib:BodekYang}%
  \BibitemOpen
  \bibfield  {author} {\bibinfo {author} {\bibfnamefont {A.}~\bibnamefont
  {Bodek}}\ and\ \bibinfo {author} {\bibfnamefont {U.~K.}\ \bibnamefont
  {Yang}},\ }\href {\doibase 10.1088/0954-3899/29/8/369} {\bibfield  {journal}
  {\bibinfo  {journal} {J. Phys. G}\ }\textbf {\bibinfo {volume} {29}},\
  \bibinfo {pages} {1899} (\bibinfo {year} {2003})}\BibitemShut {NoStop}%
\bibitem [{\citenamefont {Katori}(2015)}]{bib:MECModels}%
  \BibitemOpen
  \bibfield  {author} {\bibinfo {author} {\bibfnamefont {T.}~\bibnamefont
  {Katori}},\ }\href {\doibase 10.1063/1.4919465} {\bibfield  {journal}
  {\bibinfo  {journal} {AIP Conf. Proc.}\ }\textbf {\bibinfo {volume} {1663}},\
  \bibinfo {pages} {030001} (\bibinfo {year} {2015})}\BibitemShut {NoStop}%
\bibitem [{\citenamefont {Agostinelli}\ \emph {et~al.}(2003)\citenamefont
  {Agostinelli} \emph {et~al.}}]{bib:GEANT4}%
  \BibitemOpen
  \bibfield  {author} {\bibinfo {author} {\bibfnamefont {S.}~\bibnamefont
  {Agostinelli}} \emph {et~al.} (\bibinfo {collaboration} {GEANT4
  Collaboration}),\ }\href {\doibase 10.1016/S0168-9002(03)01368-8} {\bibfield
  {journal} {\bibinfo  {journal} {Nucl. Instrum. Meth. A}\ }\textbf {\bibinfo
  {volume} {506}},\ \bibinfo {pages} {250} (\bibinfo {year}
  {2003})}\BibitemShut {NoStop}%
\bibitem [{\citenamefont {Chou}(1952)}]{bib:BirksChou}%
  \BibitemOpen
  \bibfield  {author} {\bibinfo {author} {\bibfnamefont {C.~N.}\ \bibnamefont
  {Chou}},\ }\href {\doibase 10.1103/PhysRev.87.904} {\bibfield  {journal}
  {\bibinfo  {journal} {Phys. Rev.}\ }\textbf {\bibinfo {volume} {87}},\
  \bibinfo {pages} {904} (\bibinfo {year} {1952})}\BibitemShut {NoStop}%
\bibitem [{\citenamefont {Aurisano}\ \emph {et~al.}(2015)\citenamefont
  {Aurisano} \emph {et~al.}}]{bib:CHEP2015_sim}%
  \BibitemOpen
  \bibfield  {author} {\bibinfo {author} {\bibfnamefont {A.}~\bibnamefont
  {Aurisano}} \emph {et~al.} (\bibinfo {collaboration} {NOvA Collaboration}),\
  }\href {\doibase 10.1088/1742-6596/664/7/072002} {\bibfield  {journal}
  {\bibinfo  {journal} {J. Phys. Conf. Ser.}\ }\textbf {\bibinfo {volume}
  {664}},\ \bibinfo {pages} {072002} (\bibinfo {year} {2015})}\BibitemShut
  {NoStop}%
\bibitem [{\citenamefont {Workman}\ \emph {et~al.}(2022)\citenamefont {Workman}
  \emph {et~al.}}]{bib:PDG}%
  \BibitemOpen
  \bibfield  {author} {\bibinfo {author} {\bibfnamefont {R.}~\bibnamefont
  {Workman}} \emph {et~al.} (\bibinfo {collaboration} {Particle Data Group}),\
  }\href {\doibase 10.1093/ptep/ptac097} {\bibfield  {journal} {\bibinfo
  {journal} {Prog. Theor. Exp. Phys.}\ }\textbf {\bibinfo {volume} {C38}},\
  \bibinfo {pages} {083C01} (\bibinfo {year} {2022})}\BibitemShut {NoStop}%
\bibitem [{\citenamefont {Ester}\ \emph {et~al.}(1996)\citenamefont {Ester}
  \emph {et~al.}}]{bib:DBSCAN}%
  \BibitemOpen
  \bibfield  {author} {\bibinfo {author} {\bibfnamefont {M.}~\bibnamefont
  {Ester}} \emph {et~al.},\ }in\ \href@noop {} {\emph {\bibinfo {booktitle}
  {Proc. of 2nd International Conf. on Knowledge Discovery and Data Mining}}}\
  (\bibinfo {year} {1996})\ pp.\ \bibinfo {pages} {226--231}\BibitemShut
  {NoStop}%
\bibitem [{\citenamefont {Baird}\ \emph {et~al.}(2015)\citenamefont {Baird}
  \emph {et~al.}}]{bib:CHEP2015_reco}%
  \BibitemOpen
  \bibfield  {author} {\bibinfo {author} {\bibfnamefont {M.}~\bibnamefont
  {Baird}} \emph {et~al.},\ }\href {\doibase 10.1088/1742-6596/664/7/072035}
  {\bibfield  {journal} {\bibinfo  {journal} {J. Phys. Conf. Ser.}\ }\textbf
  {\bibinfo {volume} {664}},\ \bibinfo {pages} {072035} (\bibinfo {year}
  {2015})}\BibitemShut {NoStop}%
\bibitem [{\citenamefont {Dunn}(1973)}]{bib:FuzzyKMeans}%
  \BibitemOpen
  \bibfield  {author} {\bibinfo {author} {\bibfnamefont {J.~C.}\ \bibnamefont
  {Dunn}},\ }\href {\doibase 10.1080/01969727308546046} {\bibfield  {journal}
  {\bibinfo  {journal} {Journal of Cybernetics}\ }\textbf {\bibinfo {volume}
  {3}},\ \bibinfo {pages} {32} (\bibinfo {year} {1973})}\BibitemShut {NoStop}%
\bibitem [{\citenamefont {Kalman}(1960)}]{bib:Kalman}%
  \BibitemOpen
  \bibfield  {author} {\bibinfo {author} {\bibfnamefont {R.~E.}\ \bibnamefont
  {Kalman}},\ }\href@noop {} {\bibfield  {journal} {\bibinfo  {journal} {J.
  Basic Eng.}\ }\textbf {\bibinfo {volume} {82}} (\bibinfo {year}
  {1960})}\BibitemShut {NoStop}%
\bibitem [{\citenamefont {Raddatz}(2016)}]{bib:RaddatzThesis}%
  \BibitemOpen
  \bibfield  {author} {\bibinfo {author} {\bibfnamefont {N.~J.}\ \bibnamefont
  {Raddatz}},\ }\href {\doibase 10.2172/1253594} {Ph.D. thesis},\ \bibinfo
  {school} {U. of Minnesota} (\bibinfo {year} {2016}),\ \bibinfo {note}
  {{FERMILAB}-THESIS-2016-05}\BibitemShut {NoStop}%
\bibitem [{\citenamefont {Pershey}(2018)}]{bib:PersheyThesis}%
  \BibitemOpen
  \bibfield  {author} {\bibinfo {author} {\bibfnamefont {D.~S.}\ \bibnamefont
  {Pershey}},\ }\href {\doibase 10.2172/1484186} {Ph.D. thesis},\ \bibinfo
  {school} {Caltech} (\bibinfo {year} {2018}),\ \bibinfo {note}
  {{FERMILAB}-THESIS-2018-17}\BibitemShut {NoStop}%
\bibitem [{Note1()}]{Note1}%
  \BibitemOpen
  \bibinfo {note} {All quoted resolutions are calculated as the RMS difference
  between reconstructed and true values in simulation.}\BibitemShut {Stop}%
\bibitem [{\citenamefont {Aurisano}\ \emph {et~al.}(2016)\citenamefont
  {Aurisano} \emph {et~al.}}]{bib:CVN}%
  \BibitemOpen
  \bibfield  {author} {\bibinfo {author} {\bibfnamefont {A.}~\bibnamefont
  {Aurisano}} \emph {et~al.},\ }\href
  {http://stacks.iop.org/1748-0221/11/i=09/a=P09001} {\bibfield  {journal}
  {\bibinfo  {journal} {JINST}\ }\textbf {\bibinfo {volume} {11}},\ \bibinfo
  {pages} {P09001} (\bibinfo {year} {2016})}\BibitemShut {NoStop}%
\bibitem [{\citenamefont {D'Agostini}(1995)}]{bib:DAgostini}%
  \BibitemOpen
  \bibfield  {author} {\bibinfo {author} {\bibfnamefont {G.}~\bibnamefont
  {D'Agostini}},\ }\href {\doibase 10.1016/0168-9002(95)00274-X} {\bibfield
  {journal} {\bibinfo  {journal} {Nucl. Instrum. Meth. A}\ }\textbf {\bibinfo
  {volume} {362}},\ \bibinfo {pages} {487} (\bibinfo {year}
  {1995})}\BibitemShut {NoStop}%
\bibitem [{\citenamefont {D'Agostini}(2010)}]{bib:DAgostini2}%
  \BibitemOpen
  \bibfield  {author} {\bibinfo {author} {\bibfnamefont {G.}~\bibnamefont
  {D'Agostini}},\ }\href@noop {} {\enquote {\bibinfo {title} {Improved
  iterative bayesian unfolding},}\ } (\bibinfo {year} {2010}),\ \Eprint
  {http://arxiv.org/abs/arXiv:1010.0632} {arXiv:1010.0632} \BibitemShut
  {NoStop}%
\bibitem [{\citenamefont {Andreopoulos}\ \emph {et~al.}(2015)\citenamefont
  {Andreopoulos} \emph {et~al.}}]{bib:GENIE2}%
  \BibitemOpen
  \bibfield  {author} {\bibinfo {author} {\bibfnamefont {C.}~\bibnamefont
  {Andreopoulos}} \emph {et~al.} (\bibinfo {collaboration} {GENIE
  Collaboration}),\ }\href@noop {} {} (\bibinfo {year} {2015}),\ \Eprint
  {http://arxiv.org/abs/1510.05494} {arXiv:1510.05494 [hep-ph]} \BibitemShut
  {NoStop}%
\bibitem [{\citenamefont {Conrad}\ \emph {et~al.}(1998)\citenamefont {Conrad},
  \citenamefont {Shaevitz},\ and\ \citenamefont {Bolton}}]{RevModPhys.70.1341}%
  \BibitemOpen
  \bibfield  {author} {\bibinfo {author} {\bibfnamefont {J.~M.}\ \bibnamefont
  {Conrad}}, \bibinfo {author} {\bibfnamefont {M.~H.}\ \bibnamefont
  {Shaevitz}}, \ and\ \bibinfo {author} {\bibfnamefont {T.}~\bibnamefont
  {Bolton}},\ }\href {\doibase 10.1103/RevModPhys.70.1341} {\bibfield
  {journal} {\bibinfo  {journal} {Rev. Mod. Phys.}\ }\textbf {\bibinfo {volume}
  {70}},\ \bibinfo {pages} {1341} (\bibinfo {year} {1998})}\BibitemShut
  {NoStop}%
\bibitem [{\citenamefont {Pinzon~Guerra}\ \emph {et~al.}(2017)\citenamefont
  {Pinzon~Guerra} \emph {et~al.}}]{bib:DUETPion}%
  \BibitemOpen
  \bibfield  {author} {\bibinfo {author} {\bibfnamefont {E.~S.}\ \bibnamefont
  {Pinzon~Guerra}} \emph {et~al.} (\bibinfo {collaboration} {DUET
  Collaboration}),\ }\href {\doibase 10.1103/PhysRevC.95.045203} {\bibfield
  {journal} {\bibinfo  {journal} {Phys. Rev. C}\ }\textbf {\bibinfo {volume}
  {95}},\ \bibinfo {pages} {045203} (\bibinfo {year} {2017})}\BibitemShut
  {NoStop}%
\bibitem [{\citenamefont {Ashery}\ \emph {et~al.}(1981)\citenamefont {Ashery}
  \emph {et~al.}}]{bib:AsheryPion}%
  \BibitemOpen
  \bibfield  {author} {\bibinfo {author} {\bibfnamefont {D.}~\bibnamefont
  {Ashery}} \emph {et~al.},\ }\href {\doibase 10.1103/PhysRevC.23.2173}
  {\bibfield  {journal} {\bibinfo  {journal} {Phys. Rev. C}\ }\textbf {\bibinfo
  {volume} {23}},\ \bibinfo {pages} {2173} (\bibinfo {year}
  {1981})}\BibitemShut {NoStop}%
\bibitem [{\citenamefont {Birks}(1951)}]{bib:Birks}%
  \BibitemOpen
  \bibfield  {author} {\bibinfo {author} {\bibfnamefont {J.~B.}\ \bibnamefont
  {Birks}},\ }\href {\doibase 10.1088/0370-1298/64/10/303} {\bibfield
  {journal} {\bibinfo  {journal} {Proc. Phys. Soc.}\ }\textbf {\bibinfo
  {volume} {A64}},\ \bibinfo {pages} {874} (\bibinfo {year}
  {1951})}\BibitemShut {NoStop}%
\bibitem [{\citenamefont {Agostini}\ \emph {et~al.}(2018)\citenamefont
  {Agostini} \emph {et~al.}}]{bib:BorexinoBirks}%
  \BibitemOpen
  \bibfield  {author} {\bibinfo {author} {\bibfnamefont {M.}~\bibnamefont
  {Agostini}} \emph {et~al.} (\bibinfo {collaboration} {Borexino
  Collaboration}),\ }\href {\doibase 10.1016/j.astropartphys.2017.10.003}
  {\bibfield  {journal} {\bibinfo  {journal} {Astropart. Phys.}\ }\textbf
  {\bibinfo {volume} {97}},\ \bibinfo {pages} {136} (\bibinfo {year}
  {2018})}\BibitemShut {NoStop}%
\bibitem [{\citenamefont {Adamson}\ \emph {et~al.}(2017)\citenamefont {Adamson}
  \emph {et~al.}}]{bib:NOVA_SANumu}%
  \BibitemOpen
  \bibfield  {author} {\bibinfo {author} {\bibfnamefont {P.}~\bibnamefont
  {Adamson}} \emph {et~al.} (\bibinfo {collaboration} {NOvA Collaboration}),\
  }\href {\doibase 10.1103/PhysRevLett.118.151802} {\bibfield  {journal}
  {\bibinfo  {journal} {Phys. Rev. Lett.}\ }\textbf {\bibinfo {volume} {118}},\
  \bibinfo {pages} {151802} (\bibinfo {year} {2017})}\BibitemShut {NoStop}%
\bibitem [{\citenamefont {Formaggio}\ and\ \citenamefont
  {Zeller}(2012)}]{bib:eVtoEeV}%
  \BibitemOpen
  \bibfield  {author} {\bibinfo {author} {\bibfnamefont {J.~A.}\ \bibnamefont
  {Formaggio}}\ and\ \bibinfo {author} {\bibfnamefont {G.~P.}\ \bibnamefont
  {Zeller}},\ }\href {\doibase 10.1103/RevModPhys.84.1307} {\bibfield
  {journal} {\bibinfo  {journal} {Rev. Mod. Phys.}\ }\textbf {\bibinfo {volume}
  {84}},\ \bibinfo {pages} {1307} (\bibinfo {year} {2012})}\BibitemShut
  {NoStop}%
\bibitem [{\citenamefont {Kabirnezhad}(2018)}]{PhysRevD.97.013002}%
  \BibitemOpen
  \bibfield  {author} {\bibinfo {author} {\bibfnamefont {M.}~\bibnamefont
  {Kabirnezhad}},\ }\href {\doibase 10.1103/PhysRevD.97.013002} {\bibfield
  {journal} {\bibinfo  {journal} {Phys. Rev. D}\ }\textbf {\bibinfo {volume}
  {97}},\ \bibinfo {pages} {013002} (\bibinfo {year} {2018})}\BibitemShut
  {NoStop}%
\end{thebibliography}%
\bibliographystyle{apsrev4-1}

\onecolumngrid

\appendix

\FloatBarrier

\section{Neutrino Flux in the NuMI Beam}
\label{Appendix:Flux}

\begin{table*}[h]
\caption{The neutrino flux through the NOvA ND in bins of energy.  Histogram entries are normalized to $10^{10}$\,POT incident on the NuMI target.}

\centering
\begin{tabular}{c | c || c | c}
Energy Range [GeV] & Flux [$\nu$/cm$^2$/$10^{10}$\,POT] & Energy Range [GeV] & Flux [$\nu$/cm$^2$/$10^{10}$\,POT] \\
\hline
1.0 - 1.1 & 2.143 & 3.0 - 3.1 & 0.766 \\
1.1 - 1.2 & 2.462 & 3.1 - 3.2 & 0.601 \\
1.2 - 1.3 & 2.994 & 3.2 - 3.3 & 0.455 \\
1.3 - 1.4 & 3.896 & 3.3 - 3.4 & 0.367 \\ 
1.4 - 1.5 & 4.712 & 3.4 - 3.5 & 0.304 \\
1.5 - 1.6 & 5.405 & 3.5 - 3.6 & 0.263 \\
1.6 - 1.7 & 6.129 & 3.6 - 3.7 & 0.256 \\
1.7 - 1.8 & 6.670 & 3.7 - 3.8 & 0.225 \\
1.8 - 1.9 & 6.969 & 3.8 - 3.9 & 0.214 \\
1.9 - 2.0 & 7.050 & 3.9 - 4.0 & 0.195 \\
2.0 - 2.1 & 6.728 & 4.0 - 4.1 & 0.183 \\
2.1 - 2.2 & 6.041 & 4.1 - 4.2 & 0.177 \\
2.2 - 2.3 & 5.093 & 4.2 - 4.3 & 0.180 \\
2.3 - 2.4 & 4.060 & 4.3 - 4.4 & 0.149 \\
2.4 - 2.5 & 3.238 & 4.4 - 4.5 & 0.148 \\
2.5 - 2.6 & 2.508 & 4.5 - 4.6 & 0.153 \\
2.6 - 2.7 & 1.976 & 4.6 - 4.7 & 0.128 \\
2.7 - 2.8 & 1.554 & 4.7 - 4.8 & 0.127 \\
2.8 - 2.9 & 1.227 & 4.8 - 4.9 & 0.128 \\
2.9 - 3.0 & 0.975 & 4.9 - 5.0 & 0.127 \\
\end{tabular}

\label{table:FluxHistogram}
\end{table*}

\FloatBarrier

\section{Measured Differential Cross Section Tables}
\label{appendix:ccpi0_xsecsum}

\newcommand{\leftbar}[1]{\multicolumn{1}{|c}{#1}}

\begin{table*}
\caption[Central value and covariance matrix for CC$\pi^0$ measurement in $p_\mu$]{A summary of the extracted CC$\pi^0$ cross section, differential in $p_\mu$.  The top row and the left column give the lower edges of each analysis bin.  The second row gives the central value of the cross-section measurement in $10^{-40}\,\mathrm{cm}^2/(\mathrm{GeV}/c)$ for each kinematic bin while the third row gives the GENIE prediction.  The remaining matrix gives covariance and correlation information. Entries in the upper right are covariances in units of $10^{-80}\,\mathrm{cm}^4/(\mathrm{GeV}/c)^2$, while entries in the lower left are dimensionless correlation coefficients. All entries correspond to absolutely normalized results.  Area-normalized results that appear elsewhere in the text are presented only for those specific visualization purposes.}

\centering

\begin{small}

\begin{tabular}{ c | c c c c c c c c c c }
 & 0.00 & 0.25 & 0.50 & 0.75 & 1.00 & 1.25 & 1.50 & 1.75 & 2.00 & 3.00 \\
\hline
$\frac{d\sigma}{dp_\mu}$ & - & 27.1 & 25.2 & 21.6 & 16.2 & 12.0 & 8.68 & 5.57 & 2.13 & 0.47 \\
\hline
GENIE & 10.8 & 23.7 & 23.1 & 20.7 & 17.2 & 13.0 & 9.03 & 5.69 & 1.97 & 0.43 \\
\hline
0.00 & - & - & - & - & - & - & - & - & - & - \\ \cline{2-2}
0.25 & - & \leftbar{17.2} & 12.4 & 8.17 & 3.45 & 4.98 & 3.64 & 1.84 & 1.05 & 0.32 \\ \cline{3-3}
0.50 & - & 0.95 & \leftbar{9.96} & 6.58 & 2.83 & 4.09 & 3.14 & 1.57 & 0.95 & 0.28 \\ \cline{4-4}
0.75 & - & 0.89 & 0.94 & \leftbar{4.92} & 2.45 & 3.03 & 2.17 & 1.14 & 0.61 & 0.17 \\ \cline{5-5}
1.00 & - & 0.49 & 0.53 & 0.65 & \leftbar{2.86} & 0.85 & 0.62 & 0.44 & 0.10 & -0.03 \\ \cline{6-6}
1.25 & - & 0.72 & 0.78 & 0.82 & 0.30 & \leftbar{2.77} & 1.76 & 0.91 & 0.51 & 0.16 \\ \cline{7-7}
1.50 & - & 0.72 & 0.82 & 0.80 & 0.30 & 0.87 & \leftbar{1.48} & 0.67 & 0.42 & 0.13 \\ \cline{8-8}
1.75 & - & 0.67 & 0.75 & 0.78 & 0.39 & 0.82 & 0.82 & \leftbar{0.44} & 0.21 & 0.06 \\ \cline{9-9}
2.00 & - & 0.64 & 0.76 & 0.69 & 0.15 & 0.78 & 0.88 & 0.80 & \leftbar{0.15} & 0.05 \\ \cline{10-10}
3.00 & - & 0.45 & 0.52 & 0.46 & -0.11 & 0.58 & 0.61 & 0.53 & 0.68 & \leftbar{0.03} \\
\end{tabular}

\end{small}

\label{table:ccpi0_xseccov_pmu}
\end{table*}

\begin{table*}

\caption{Same as in Table~\ref{table:ccpi0_xseccov_pmu}, but for $\cos\theta_\mu$ with cross-section units of $10^{-40}\,\mathrm{cm}^2$ and covariance units of $10^{-80}\,\mathrm{cm}^4$.}

\centering

\begin{small}

\begin{tabular}{ c | c c c c c c c c c c c }
 & -1.00 & -0.50 & 0.00 & 0.25 & 0.50 & 0.60 & 0.70 & 0.80 & 0.85 & 0.90 & 0.95 \\
\hline
$\frac{d\sigma}{d\cos\theta_\mu}$ & 1.72 & 3.49 & 6.64 & 11.9 & 19.8 & 27.9 & 40.8 & 59.6 & 79.5 & 107. & 136. \\
\hline
GENIE & 1.63 & 3.29 & 6.25 & 11.1 & 17.9 & 25.4 & 38.5 & 55.5 & 74.2 & 103. & 132. \\
\hline
-1.00 & 0.16 & 0.25 & 0.46 & 0.77 & 0.95 & 1.61 & 1.78 & 2.11 & 2.95 & 3.92 & 3.51 \\ \cline{2-2}
-0.50 & 0.97 & \leftbar{0.42} & 0.74 & 1.19 & 1.45 & 2.46 & 2.93 & 3.52 & 4.83 & 6.35 & 5.71 \\ \cline{3-3}
0.00 & 0.93 & 0.93 & \leftbar{1.51} & 2.92 & 3.32 & 5.77 & 6.14 & 7.00 & 9.95 & 13.1 & 12.4 \\ \cline{4-4}
0.25 & 0.73 & 0.70 & 0.90 & \leftbar{7.03} & 7.19 & 13.1 & 11.3 & 12.4 & 18.9 & 25.3 & 25.4 \\ \cline{5-5}
0.50 & 0.76 & 0.72 & 0.87 & 0.87 & \leftbar{9.73} & 14.5 & 15.2 & 17.6 & 25.8 & 34.7 & 36.9 \\ \cline{6-6}
0.60 & 0.79 & 0.75 & 0.93 & 0.97 & 0.92 & \leftbar{25.8} & 24.0 & 27.2 & 40.6 & 54.1 & 55.8 \\ \cline{7-7}
0.70 & 0.77 & 0.78 & 0.87 & 0.74 & 0.85 & 0.82 & \leftbar{33.4} & 37.1 & 50.1 & 61.8 & 66.9 \\ \cline{8-8}
0.80 & 0.78 & 0.80 & 0.84 & 0.69 & 0.83 & 0.79 & 0.95 & \leftbar{45.9} & 59.8 & 75.4 & 82.5 \\ \cline{9-9}
0.85 & 0.79 & 0.80 & 0.87 & 0.77 & 0.89 & 0.86 & 0.93 & 0.95 & \leftbar{86.4} & 111. & 123. \\ \cline{10-10}
0.90 & 0.77 & 0.77 & 0.84 & 0.75 & 0.88 & 0.84 & 0.84 & 0.88 & 0.94 & \leftbar{161.} & 160. \\ \cline{11-11}
0.95 & 0.61 & 0.62 & 0.71 & 0.67 & 0.83 & 0.77 & 0.81 & 0.85 & 0.93 & 0.88 & \leftbar{203.} \\
\end{tabular}

\end{small}

\label{table:ccpi0_xseccov_cosmu}
\end{table*}

\begin{table*}

\caption{Same as in Table~\ref{table:ccpi0_xseccov_pmu}, but for $p_\pi$.}

\centering

\begin{small}

\begin{tabular}{ c | c c c c c c c c c c c c c c c c }
 & 0.00 & 0.10 & 0.20 & 0.30 & 0.40 & 0.50 & 0.60 & 0.70 & 0.80 & 0.90 & 1.00 & 1.25 & 1.50 & 1.75 & 2.00 & 2.50 \\
\hline
$\frac{d\sigma}{dp_\pi}$ & 20.3 & 55.1 & 48.6 & 46.2 & 40.6 & 32.0 & 25.1 & 19.3 & 15.3 & 11.2 & 7.39 & 3.95 & 2.07 & 1.05 & 0.41 & 0.12 \\
\hline
GENIE & 19.2 & 53.3 & 48.1 & 45.5 & 38.9 & 29.8 & 22.6 & 17.0 & 13.3 & 9.75 & 6.48 & 3.50 & 1.85 & 0.97 & 0.40 & 0.12 \\
\hline

0.00 & 11.3 & 22.0 & 17.0 & 14.2 & 10.6 & 7.59 & 5.76 & 3.99 & 2.69 & 1.71 & 0.90 & 0.41 & 0.19 & 0.70 & -0.013 & -0.029 \\ \cline{2-2}
0.10 & 0.80 & \leftbar{67.6} & 59.2 & 49.5 & 35.0 & 21.8 & 16.9 & 10.7 & 6.78 & 4.04 & 1.77 & 0.82 & 0.44 & -0.008 & -0.15 & -0.15 \\ \cline{3-3}
0.20 & 0.68 & 0.96 & \leftbar{55.8} & 46.8 & 32.5 & 19.4 & 15.3 & 9.44 & 5.83 & 3.63 & 1.56 & 0.83 & 0.55 & 0.08 & -0.12 & -0.138 \\ \cline{4-4}
0.30 & 0.65 & 0.93 & 0.96 & \leftbar{42.3} & 30.1 & 18.4 & 14.5 & 9.11 & 5.80 & 3.75 & 1.85 & 1.10 & 0.70 & 0.23 & -0.05 & -0.09 \\ \cline{5-5}
0.40 & 0.65 & 0.88 & 0.90 & 0.95 & \leftbar{23.6} & 15.2 & 11.9 & 7.63 & 5.25 & 3.19 & 1.81 & 1.06 & 0.63 & 0.23 & 0.02 & -0.04 \\ \cline{6-6}
0.50 & 0.68 & 0.79 & 0.78 & 0.85 & 0.94 & \leftbar{11.1} & 8.22 & 5.59 & 4.11 & 2.41 & 1.52 & 0.84 & 0.48 & 0.19 & 0.05 & -0.02 \\ \cline{7-7}
0.60 & 0.67 & 0.80 & 0.80 & 0.87 & 0.95 & 0.96 & \leftbar{6.62} & 4.32 & 3.13 & 1.95 & 1.23 & 0.70 & 0.37 & 0.17 & 0.05 & -0.005 \\ \cline{8-8}
0.70 & 0.68 & 0.74 & 0.72 & 0.80 & 0.90 & 0.96 & 0.96 & \leftbar{3.06} & 2.25 & 1.42 & 0.93 & 0.50 & 0.26 & 0.11 & 0.04 & -0.005 \\ \cline{9-9}
0.80 & 0.59 & 0.61 & 0.58 & 0.66 & 0.80 & 0.91 & 0.90 & 0.95 & \leftbar{1.83} & 1.09 & 0.78 & 0.43 & 0.23 & 0.11 & 0.04 & 0.004 \\ \cline{10-10}
0.90 & 0.56 & 0.54 & 0.54 & 0.64 & 0.72 & 0.80 & 0.84 & 0.90 & 0.89 & \leftbar{0.82} & 0.54 & 0.306 & 0.16 & 0.09 & 0.03 & 0.005 \\ \cline{11-11}
1.00 & 0.42 & 0.33 & 0.33 & 0.44 & 0.58 & 0.71 & 0.74 & 0.82 & 0.89 & 0.93 & \leftbar{0.41} & 0.24 & 0.13 & 0.07 & 0.03 & 0.009 \\ \cline{12-12}
1.25 & 0.30 & 0.24 & 0.27 & 0.41 & 0.53 & 0.61 & 0.66 & 0.70 & 0.76 & 0.81 & 0.91 & \leftbar{0.17} & 0.10 & 0.06 & 0.02 & 0.009 \\ \cline{13-13}
1.50 & 0.21 & 0.20 & 0.27 & 0.39 & 0.47 & 0.53 & 0.53 & 0.54 & 0.62 & 0.65 & 0.74 & 0.85 & \leftbar{0.08} & 0.04 & 0.01 & 0.005 \\ \cline{14-14}
1.75 & 0.12 & -0.01 & 0.06 & 0.19 & 0.27 & 0.32 & 0.37 & 0.36 & 0.44 & 0.56 & 0.65 & 0.78 & 0.82 & \leftbar{0.03} & 0.01 & 0.005 \\ \cline{15-15}
2.00 & -0.05 & -0.24 & -0.23 & -0.10 & 0.04 & 0.20 & 0.24 & 0.28 & 0.42 & 0.44 & 0.65 & 0.72 & 0.61 & 0.77 & \leftbar{0.005} & 0.003 \\ \cline{16-16}
2.50 & -0.16 & -0.36 & -0.35 & -0.27 & -0.16 & -0.09 & -0.04 & -0.06 & 0.06 & 0.11 & 0.28 & 0.42 & 0.35 & 0.57 & 0.72 & \leftbar{0.003} \\

\end{tabular}

\end{small}

\label{table:ccpi0_xseccov_ppi}
\end{table*}

\begin{table*}
\caption{Same as in Table~\ref{table:ccpi0_xseccov_pmu}, but for $\cos\theta_\pi$ with cross-section units of $10^{-40}\,\mathrm{cm}^2$ and covariance units of $10^{-80}\,\mathrm{cm}^4$.}

\centering

\begin{small}

\begin{tabular}{ c | c c c c c c c c c c c c c c }
 & -1.00 & -0.75 & -0.50 & -0.25 & 0.00 & 0.10 & 0.20 & 0.30 & 0.40 & 0.50 & 0.60 & 0.70 & 0.80 & 0.90 \\
\hline
$\frac{d\sigma}{d\cos\theta_\pi}$ & 4.73 & 5.32 & 6.39 & 8.12 & 10.0 & 11.5 & 13.2 & 15.2 & 17.8 & 21.5 & 27.1 & 35.6 & 51.3 & 92.5 \\
\hline
GENIE & 4.45 & 4.93 & 5.77 & 7.17 & 8.66 & 9.76 & 11.2 & 13.2 & 15.7 & 19.2 & 24.4 & 32.7 & 4.86 & 93.1 \\
\hline
-1.00 & 0.48 & 0.57 & 0.67 & 0.74 & 1.05 & 0.94 & 1.13 & 1.23 & 1.04 & 1.54 & 2.09 & 2.49 & 4.07 & 7.56 \\ \cline{2-2}
-0.75 & 0.97 & \leftbar{0.73} & 0.82 & 1.88 & 1.31 & 1.45 & 1.41 & 1.57 & 1.29 & 1.93 & 2.65 & 3.03 & 4.92 & 9.27 \\ \cline{3-3}
-0.50 & 0.96 & 0.95 & \leftbar{1.01} & 1.03 & 1.56 & 1.35 & 1.66 & 1.88 & 1.48 & 2.27 & 3.18 & 3.59 & 5.85 & 11.2 \\ \cline{4-4}
-0.25 & 0.98 & 0.94 & 0.94 & \leftbar{1.20} & 1.66 & 1.51 & 1.76 & 1.93 & 1.66 & 2.45 & 3.25 & 3.92 & 6.28 & 11.5 \\ \cline{5-5}
0.00 & 0.92 & 0.93 & 0.95 & 0.93 & \leftbar{2.69} & 2.20 & 2.61 & 3.07 & 2.29 & 3.69 & 5.02 & 5.52 & 8.84 & 17.0 \\ \cline{6-6}
0.10 & 0.95 & 0.95 & 0.95 & 0.97 & 0.95 & \leftbar{2.01} & 2.35 & 2.70 & 2.23 & 3.37 & 4.47 & 5.18 & 8.11 & 15.3 \\ \cline{7-7}
0.20 & 0.95 & 0.96 & 0.96 & 0.93 & 0.93 & 0.97 & \leftbar{2.95} & 3.29 & 2.76 & 4.08 & 5.60 & 6.41 & 10.2 & 19.4 \\ \cline{8-8}
0.30 & 0.87 & 0.91 & 0.92 & 0.87 & 0.92 & 0.94 & 0.94 & \leftbar{4.11} & 3.09 & 4.89 & 6.63 & 7.12 & 10.9 & 21.7 \\ \cline{9-9}
0.40 & 0.87 & 0.87 & 0.85 & 0.87 & 0.80 & 0.91 & 0.93 & 0.88 & \leftbar{3.02} & 4.03 & 5.35 & 6.42 & 9.80 & 18.42 \\ \cline{10-10}
0.50 & 0.88 & 0.90 & 0.90 & 0.89 & 0.90 & 0.95 & 0.95 & 0.96 & 0.92 & \leftbar{6.30} & 8.17 & 9.23 & 14.0 & 27.3 \\ \cline{11-11}
0.60 & 0.90 & 0.92 & 0.94 & 0.88 & 0.91 & 0.94 & 0.97 & 0.97 & 0.92 & 0.97 & \leftbar{11.3} & 12.4 & 19.3 & 38.0 \\ \cline{12-12}
0.70 & 0.92 & 0.91 & 0.92 & 0.92 & 0.86 & 0.94 & 0.96 & 0.90 & 0.95 & 0.95 & 0.95 & \leftbar{15.1} & 23.4 & 44.2 \\ \cline{13-13}
0.80 & 0.95 & 0.93 & 0.94 & 0.93 & 0.87 & 0.93 & 0.96 & 0.87 & 0.91 & 0.90 & 0.93 & 0.98 & \leftbar{38.1} & 71.0 \\ \cline{14-14}
0.90 & 0.93 & 0.92 & 0.95 & 0.90 & 0.88 & 0.92 & 0.96 & 0.91 & 0.90 & 0.93 & 0.96 & 0.97 & 0.98 & \leftbar{138.} \\
\end{tabular}

\end{small}

\label{table:ccpi0_xseccov_cospi}
\end{table*}

\begin{table*}
\caption{Same as in Table~\ref{table:ccpi0_xseccov_pmu}, but for $Q^2$ with cross-section units of $10^{-40}\,\mathrm{cm}^2/(\mathrm{GeV}/c)^2$ and covariance units of $10^{-80}\,\mathrm{cm}^4/(\mathrm{GeV}/c)^4$.}

\centering

\begin{small}

\begin{tabular}{ c | c c c c c c c c c c c }
 & 0.00 & 0.10 & 0.25 & 0.50 & 0.75 & 1.00 & 1.25 & 1.50 & 1.75 & 2.00 & 3.00 \\
\hline
$\frac{d\sigma}{dQ^2}$ & 16.9 & 28.7 & 30.9 & 25.3 & 19.2 & 14.0 & 9.76 & 6.54 & 4.28 & 1.68 & 0.35 \\
\hline
GENIE & 17.1 & 26.8 & 29.1 & 24.4 & 18.4 & 12.7 & 8.66 & 5.77 & 3.77 & 1.46 & 0.31 \\
\hline
0.00 & 5.03 & 6.58 & 6.77 & 5.11 & 4.33 & 3.27 & 2.07 & 1.31 & 0.69 & 0.30 & -0.008 \\ \cline{2-2}
0.10 & 0.82 & \leftbar{12.8} & 11.3 & 8.57 & 8.16 & 5.52 & 3.05 & 1.89 & 1.02 & 0.46 & 0.002 \\ \cline{3-3}
0.25 & 0.85 & 0.88 & \leftbar{12.7} & 11.4 & 9.67 & 7.18 & 4.81 & 3.53 & 2.27 & 0.91 & 0.08 \\ \cline{4-4}
0.50 & 0.66 & 0.69 & 0.92 & \leftbar{12.0} & 9.56 & 7.35 & 5.31 & 4.23 & 2.90 & 1.13 & 0.14 \\ \cline{5-5}
0.75 & 0.66 & 0.78 & 0.92 & 0.94 & \leftbar{8.61} & 6.22 & 4.27 & 3.35 & 2.18 & 0.88 & 0.09 \\ \cline{6-6}
1.00 & 0.66 & 0.69 & 0.91 & 0.96 & 0.96 & \leftbar{4.93} & 3.42 & 2.77 & 1.79 & 0.72 & 0.07 \\ \cline{7-7}
1.25 & 0.56 & 0.52 & 0.82 & 0.94 & 0.89 & 0.94 & \leftbar{2.67} & 2.15 & 1.47 & 0.58 & 0.07 \\ \cline{8-8}
1.50 & 0.43 & 0.39 & 0.72 & 0.89 & 0.83 & 0.91 & 0.96 & \leftbar{1.88} & 1.26 & 0.49 & 0.06 \\ \cline{9-9}
1.75 & 0.31 & 0.29 & 0.65 & 0.86 & 0.76 & 0.83 & 0.92 & 0.94 & \leftbar{0.95} & 0.35 & 0.05 \\ \cline{10-10}
2.00 & 0.35 & 0.34 & 0.69 & 0.87 & 0.81 & 0.87 & 0.95 & 0.96 & 0.97 & \leftbar{0.14} & 0.02 \\ \cline{11-11}
3.00 & -0.05 & 0.01 & 0.32 & 0.56 & 0.43 & 0.47 & 0.59 & 0.63 & 0.74 & 0.72 & \leftbar{0.005} \\
\end{tabular}

\end{small}

\label{table:ccpi0_xseccov_q2}
\end{table*}

\begin{table*}
\caption{Same as in Table~\ref{table:ccpi0_xseccov_pmu}, but for $W$ with cross-section units of $10^{-40}\,\mathrm{cm}^2/(\mathrm{GeV}/c^2)$ and covariance units of $10^{-80}\,\mathrm{cm}^4/(\mathrm{GeV}/c^2)^2$.}

\centering

\begin{small}

\begin{tabular}{ c | c c c c c c c c c c c c }
 & 1.00 & 1.10 & 1.20 & 1.30 & 1.40 & 1.50 & 1.60 & 1.70 & 1.80 & 2.00 & 2.25 & 2.50 \\
\hline
$\frac{d\sigma}{dW}$ & 6.11 & 20.4 & 35.2 & 39.2 & 43.5 & 48.2 & 46.8 & 37.8 & 25.1 & 10.7 & 3.81 & 0.95 \\
\hline
GENIE & 5.26 & 18.3 & 33.0 & 34.4 & 37.4 & 43.1 & 43.8 & 35.0 & 21.9 & 8.94 & 3.04 & 0.68 \\
\hline
1.00 & 3.88 & 9.39 & 6.68 & 5.53 & 6.06 & 7.06 & 9.86 & 7.59 & 2.98 & 1.41 & -0.35 & -0.49 \\ \cline{2-2}
1.10 & 0.93 & \leftbar{26.3} & 19.5 & 13.3 & 14.5 & 17.7 & 27.1 & 21.9 & 9.92 & 5.05 & -0.06 & -0.72 \\ \cline{3-3}
1.20 & 0.73 & 0.82 & \leftbar{21.6} & 16.8 & 15.1 & 16.6 & 21.5 & 15.8 & 8.24 & 5.49 & 1.07 & -0.01 \\ \cline{4-4}
1.30 & 0.63 & 0.58 & 0.81 & \leftbar{20.0} & 20.0 & 21.2 & 19.5 & 13.7 & 6.21 & 2.46 & -0.43 & -0.88 \\ \cline{5-5}
1.40 & 0.60 & 0.55 & 0.63 & 0.87 & \leftbar{26.4} & 29.3 & 25.8 & 19.5 & 8.05 & 0.90 & -1.96 & -1.85 \\ \cline{6-6}
1.50 & 0.60 & 0.58 & 0.59 & 0.79 & 0.95 & \leftbar{36.2} & 32.3 & 24.4 & 9.91 & 1.40 & -2.36 & -2.34 \\ \cline{7-7}
1.60 & 0.81 & 0.86 & 0.75 & 0.71 & 0.81 & 0.87 & \leftbar{37.9} & 30.3 & 13.7 & 4.18 & -1.14 & -1.62 \\ \cline{8-8}
1.70 & 0.75 & 0.83 & 0.66 & 0.59 & 0.74 & 0.79 & 0.95 & \leftbar{26.6} & 12.5 & 3.32 & -0.72 & -1.08 \\ \cline{9-9}
1.80 & 0.55 & 0.71 & 0.65 & 0.51 & 0.57 & 0.60 & 0.81 & 0.89 & \leftbar{7.52} & 2.29 & 0.62 & 0.11 \\ \cline{10-10}
2.00 & 0.41 & 0.56 & 0.68 & 0.32 & 0.10 & 0.13 & 0.39 & 0.37 & 0.48 & \leftbar{3.04} & 1.26 & 0.54 \\ \cline{11-11}
2.25 & -0.17 & -0.01 & 0.21 & -0.09 & -0.35 & -0.36 & -0.17 & -0.13 & 0.21 & 0.67 & \leftbar{1.16} & 0.68 \\ \cline{12-12}
2.50 & -0.36 & -0.20 & -0.00 & -0.29 & -0.52 & -0.57 & -0.38 & -0.30 & 0.06 & 0.45 & 0.91 & \leftbar{0.47} \\

\end{tabular}

\end{small}

\label{table:ccpi0_xseccov_W}
\end{table*}

\end{document}